\documentclass[%
 reprint,
 nofootinbib,
 amsmath,amssymb,
 aps,
 prd,
 longbibliography,
 nobalancelastpage,
]{revtex4-2}
\usepackage{graphicx}% Include figure files
\usepackage{dcolumn}% Align table columns on decimal point
\usepackage{bm}% bold math
\usepackage{hyperref}% add hypertext capabilities
\usepackage{comment}
\usepackage{orcidlink}

\usepackage{physics}

\usepackage[normalem]{ulem} % 横線を引くためのパッケージ

\usepackage{aas_macros}

\usepackage{subfigure}

\usepackage{booktabs} % 罫線のスタイルを変更するパッケージ
\usepackage{siunitx}  % 単位の整形に使うパッケージ
\usepackage{caption} % キャプションの設定に使うパッケージ
\usepackage{enumitem}

\usepackage{multirow} 

\usepackage{tikz}
\usepackage{xcolor}
\definecolor{RED}{rgb}{1,0,0} % 色を定義する

\usepackage{ragged2e} % パッケージの追加

\usepackage{ulem} % 横線を引くためのパッケージ

\usepackage[utf8]{inputenc}
\newcommand{\ioka}[1]{\textcolor{black}{#1}}

\newcommand{\niShiura}[1]{\textcolor{black}{#1}}

\newcommand{\Rei}[1]{}

\begin{document}
%\linenumbers % 行番号を有効にする

%\preprint{APS/123-QED}

\title{Induced Scattering of Fast Radio Bursts in Magnetar Magnetospheres}% Force line breaks with \\

\author{Rei Nishiura\orcidlink{0009-0003-8209-5030}}
 \email{nishiura@tap.scphys.kyoto-u.ac.jp}
 \affiliation{%
 Department of Physics, Kyoto University, Kyoto 606-8502, Japan}%
 \author{Shoma F. Kamijima\orcidlink{0000-0002-4821-170X}}%
 \email{shoma.kamijima@yukawa.kyoto-u.ac.jp}
\affiliation{%
 Center for Gravitational Physics and Quantum Information, 
 Yukawa Institute for Theoretical Physics, Kyoto University, Kyoto 606-8502, Japan}%
 \author{Kunihito Ioka\orcidlink{0000-0002-3517-1956}}%
 \email{kunihito.ioka@yukawa.kyoto-u.ac.jp}
\affiliation{%
 Center for Gravitational Physics and Quantum Information, 
 Yukawa Institute for Theoretical Physics, Kyoto University, Kyoto 606-8502, Japan}%

%\affiliation{
%\affiliation{
%\affiliation{%
%\selectlanguage{english}
\date{\today}% It is always \today, today,

\begin{abstract}
We investigate induced Compton/Brillouin scattering of electromagnetic waves in magnetized electron and positron pair plasma by verifying kinetic theory with Particle-in-Cell simulations. Applying this to fast radio bursts (FRBs) in magnetar magnetospheres, we find that the scattering--although suppressed by the magnetic field--inevitably enters the linear growth stage before \ioka{the incident wave amplitude becomes comparable to the background magnetic field during its outward propagation through the magnetosphere.} The subsequent evolution bifurcates: full scattering occurs when the density exceeds a critical value, whereas below it the scattering saturates and the FRB can escape without substantial induced-scattering attenuation. This nonlinear saturation eases the tension with observations of compact emission regions and may explain the observed diversity, including the presence or absence of FRBs associated with X-ray bursts.
\end{abstract}

\maketitle

%\tableofcontents
\section{Introduction}
Fast radio bursts (FRBs) are the brightest radio transients, first discovered in 2007 \citep{2007Sci...318..777L}. Most FRBs are extragalactic, and their progenitors and radiation mechanisms remain unresolved. The simultaneous detection in 2020 of FRB 20200428 with X-ray bursts from a Galactic magnetar established magnetars as at least one class of FRB sources \citep{2020Natur.587...54C,2020Natur.587...59B,2020ApJ...898L..29M}.

\begin{figure*}
  \centering
  \includegraphics[width=\textwidth]{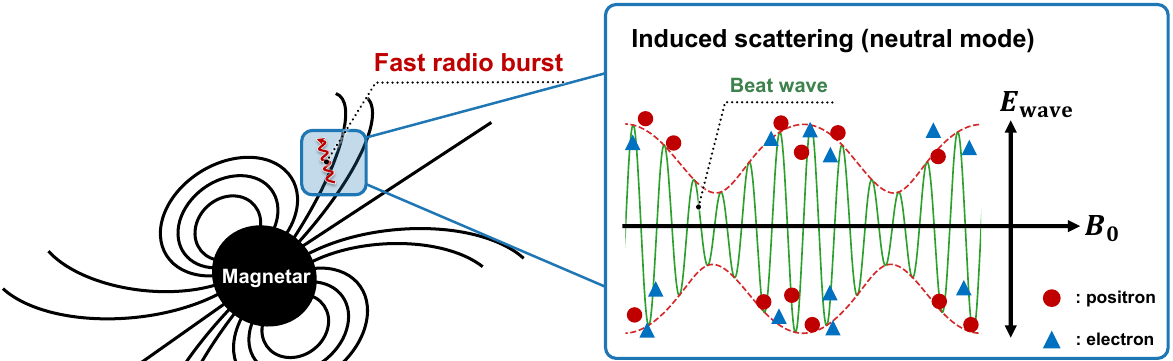}
  \caption{\justifying Schematic illustration of neutral-mode induced scattering of an FRB in a magnetar magnetosphere.}
  \label{fig:induced_scattering_from_magnetar}
\end{figure*}

Nonlinear interactions between large-amplitude electromagnetic (EM) waves and plasma are central in both astrophysical and laboratory environments, such as the Sun~\citep{1963SPhD....7..988G,1966PhFl....9.1483B,1972JPlPh...8..197B,1978ApJ...224.1013D,1990JGR....9510525I,1993JGR....9813247J,1994JGR....9923431H,2001A&A...367..705D,2006PhPl...13l4501N,2015JPlPh..81a3202D,2017ApJ...842...63S,2022RvMPP...6...22N}, FRBs \citep{2008ApJ...682.1443L,2023MNRAS.522.2133I,2024PhRvE.110a5205I}, pulsars \citep{1973PhFl...16.1480M,1975Ap&SS..36..303B,1976MNRAS.174...59B,1978MNRAS.185..297W,1982MNRAS.200..881W,1996AstL...22..399L}, laser–plasma interactions \citep{1973PhFl...16.1522K,1974PhRvL..33..209M,1975PhFl...18.1002F,1994PhPl....1.1626T,1996PhRvL..77.2483D} and free electron lasers \citep{1979PhFl...22.1089K,1980PhFl...23.2376F}. Parametric instability redistributes the incident wave energy into scattered waves and plasma through the beat of these waves, as illustrated in Fig.~\ref{fig:induced_scattering_from_magnetar}. Among such processes, we collectively refer to induced Compton scattering (ICS), stimulated Brillouin scattering (SBS), and stimulated Raman scattering (SRS) as \emph{induced scattering}. ICS denotes coupling between the beat and particle thermal motion via Landau resonance, whereas SBS and SRS are three-wave resonances, in which the beat couples to an acoustic wave and a Langmuir wave, respectively.

A key theoretical concern for magnetar FRBs is that induced scattering may attenuate emission within the magnetosphere so strongly that the radiation cannot escape \citep{2021ApJ...922L...7B,2022PhRvL.128y5003B,2022MNRAS.515.2020Q,2022arXiv221013506C,2023ApJ...959...34B,2024ApJ...975..223B,2024ApJ...975..226H,2024PhRvD.109d3048N,2024A&A...690A.332S,2024PhRvE.110a5205I,2025PhRvD.111f3055N}. On the other hand, several observations imply that the source size of FRBs is comparable to the magnetosphere~\citep{2022Natur.607..256C,2022NatAs...6..393N,2025Natur.637...48N}, leaving tension between theoretical expectations and observation.

Despite this importance, induced scattering in strongly magnetized plasma is not fully understood \citep{1978A&A....66..139S,1998PhRvE..57..994M,1999PhRvE..59.4552M,2003PhRvE..67d6406M,2006EP&S...58.1213M,2012PhPl...19h2104L,2014PhPl...21c2102L,2014NPGeo..21..217M,2024PhRvE.110a5205I}. \citet{2025PhRvD.111f3055N,2025arXiv251012869N} first developed a kinetic framework of induced scattering in strongly magnetized $e^\pm$ plasma and identified three density fluctuation modes, \emph{the ordinary, neutral, and charged modes}. They also showed that ICS, SBS, and SRS can compete with one another.

In the follow-up work, \citet{2026arXiv260101169K} used a one-dimensional Particle-in-Cell (PIC) simulation of a circularly polarized Alfvén wave. This simulation clarified two key points. First, it confirmed the linear growth rates of neutral ICS and charged ICS predicted by the analytical works \citep{2025PhRvD.111f3055N,2025arXiv251012869N}. It also showed that lowering the plasma temperature suppresses the Debye-screened charged ICS and makes neutral ICS dominant. Second, it followed the nonlinear evolution beyond the linear growth stage and showed that, in some cases, the instability saturates while the incident wave energy is hardly damped.

In this work, we identify the regime in which induced scattering prevents FRBs from propagating through a magnetar magnetosphere. In the linear growth stage, we test the transition from ICS to SBS in the neutral mode predicted by the analytical work \citep{2025arXiv251012869N}. In the nonlinear stage, we study the regime in which the incident wave is attenuated before the instability saturates. We then apply these linear and nonlinear results to induced scattering of FRBs in magnetar magnetospheres. This work is complementary to \citet{2026arXiv260101169K}, who examined parameter regions in which nonlinear saturation allows FRBs to escape with little attenuation by induced scattering.

\section{Summary of the kinetic theory of induced scattering}
In this section, we summarize the kinetic theory of induced scattering for an incident EM wave whose electric field $\bm{E}_{\mathrm{w}0}$ is perpendicular to the strong background magnetic field $\bm{B}_0$ in the high density regime $\omega_0\sim\omega_1 \ll \omega_{\mathrm p} \ll \omega_{\mathrm c}$, that is appropriate for magnetar magnetospheres \citep{2025PhRvD.111f3055N,2025arXiv251012869N}. Here $\omega_0$ and $\omega_1$ are the frequencies of the incident and scattered waves, $\omega_{\mathrm c}\equiv eB_0/(m_{\mathrm e}c)$ is the cyclotron frequency, and $\omega_{\mathrm p}\equiv \sqrt{8\pi e^2 n_{\mathrm{e}0}/m_{\mathrm e}}$ is the plasma frequency. The polarization $\bm{E}_{\mathrm{w}0}\perp\bm{B}_0$ corresponds to a fast magnetosonic wave that propagates in an arbitrary direction or to an Alfvén wave with wave vector $\bm{k}_0\parallel\bm{B}_0$. In contrast, an EM pulse with $\bm{E}_{\mathrm{w}0}\parallel \bm{B}_0$ does not propagate because of the cutoff \citep{2025PhRvD.111f3055N,2025arXiv251012869N}. We assume that the motion of the charged particles in the plasma is nonrelativistic and can be characterized by
\begin{equation}
\eta_{\mathrm{inci}} \equiv \frac{B_{\mathrm{inci}}}{B_0}
= a_{\mathrm e}\,\frac{\omega_{0}}{\omega_{\mathrm c}}
\left(1+\frac{\omega_{\mathrm p}^2}{\omega_{\mathrm c}^2}\right)^{\frac{1}{2}}\ll1,
\label{eq:definition_of_eta_letter}
\end{equation}
where $|\bm{B}_{\mathrm{w}0}|\equiv B_{\mathrm{inci}}$ is the incident magnetic amplitude and $a_{\mathrm e}\equiv eE_{\mathrm{inci}}/(m_{\mathrm e}c\,\omega_0)$ is the strength parameter. 

We refer to \citep{2025arXiv251012869N} for details of the kinetic framework and summarize here only the key elements. The framework is based on the Vlasov equation for distibution functions $f_{\pm}(\bm{r}, \bm{v}, t)=f_0(\bm{v})+\delta f_{\pm}(\bm{r}, \bm{v}, t)$ including the ponderomotive potential generated by the beat wave,
\begin{equation}
\frac{\partial f_{\pm}}{\partial t} + \bm{v} \cdot \bm{\nabla} f_{\pm} + \left[-\bm{\nabla} \phi^{\pm}_{\mathrm{p}} \pm e \left( \bm{E} + \frac{\bm{v} \times \bm{B}_{0}}{c} \right)\right] \cdot \frac{\partial f_{\pm}}{\partial \bm{p}} = 0,
\label{eq:Vlasov_eq_letter}
\end{equation}
and is solved together with the Maxwell equations $\bm{\nabla} \cdot \bm{E} = \sum_{q = \pm e} 4 \pi q n_{\text{e} 0} \int \delta f_{\pm}\,\dd^{3} \bm{v}$. Here $+~(-)$ denotes a positron (electron), and the equilibrium distribution $f_0(\bm{v})$ is assumed to be Maxwellian. The magnetized ponderomotive potential includes one $\bm{E}_{\mathrm{w}}\parallel\bm{B}_0$ term and two $\bm{E}_{\mathrm{w}}\perp\bm{B}_0$ terms \citep{1968CzJPh..18.1280K,1977PhRvL..39..402C,1981PhFl...24.1238C,1981PhRvL..46..240H,10.1063/1.864196,1996GeoRL..23..327L}. Here we restrict the propagating EM waves to $\bm{E}_{\mathrm{w}}\perp\bm{B}_0$, so that we neglect the $\bm{E}_{\mathrm{w}}\parallel\bm{B}_0$ term. The remaining transverse ponderomotive potential is expressed as
\begin{equation}
\begin{aligned}
\phi^{\pm}_{\mathrm{p}}
= \frac{e^{2}}{2 m_{\text{e}}}
\Biggl\langle
 - \frac{|\bm{E}_{\mathrm{w}\perp}|^{2}}{\omega_{\text{c}}^{2} - \omega_{0}^{2}}
 \pm \mathrm{i}\,\frac{\omega_{\text{c}}\,
  \hat{\bm{B}}_{0}\!\cdot\!\big(\bm{E}_{\mathrm{w}\perp}^{*}\!\times\!\bm{E}_{\mathrm{w}\perp}\big)}
  {\omega_{0}\big(\omega_{0}^{2}-\omega_{\text{c}}^{2} \big)}
\Biggr\rangle,
\end{aligned}
\label{eq:ponderomotive_potential_letter}
\end{equation}
where $\omega_{1} \sim \omega_{0} \gg |\omega|$, the time average is taken over a timescale longer than $\omega_0^{-1}$ and shorter than $|\omega|^{-1}$, and $\omega\equiv\omega_{1}-\omega_{0}$ is the density fluctuation frequency. The first term in Eq.~\eqref{eq:ponderomotive_potential_letter} drives the neutral mode and the second term drives the charged mode. As illustrated in Fig.~\ref{fig:induced_scattering_from_magnetar}, the neutral mode excites a density fluctuation independent of the charge sign, similar to the unmagnetized case, whereas the charged mode excites a density fluctuation accompanied by charge separation, which is distinct from the unmagnetized case. 

Linearizing Eq.~\eqref{eq:Vlasov_eq_letter} yields the dispersion relation and the linear growth rate of the scattered wave and the density fluctuation, which we define as $t^{-1}\equiv 2~\text{Im}~\omega_{1}=2~\text{Im}~\omega$. The neutral mode exhibits a continuous transition from ICS to SBS as the incident amplitude $a_{\mathrm e}\omega_0/\omega_{\mathrm c}$ increases. $(t_{\mathrm{neutral}}^{\mathrm{coh}})^{-1}\propto a_{\mathrm e}^{2}$ in the ICS regime and $(t_{\mathrm{neutral}}^{\mathrm{coh}})^{-1}\propto a_{\mathrm e}^{\frac{2}{3}}$ in the SBS regime. The precise expression for the maximum linear growth rate, together with the transition amplitude $\left(a_{\mathrm{e}}\omega_{0}/\omega_{\mathrm c}\right)_{\mathrm{trans}}^{\mathrm{coh}}$, is summarized in Eqs.~\eqref{eq:growth_rate_neutral_matome_Brillouin_letter} and \eqref{eq:transition_point_for_neutral_mode_letter}. The maximum growth occurs when an EM wave incident along $\bm{B}_0$ that undergoes $180^\circ$ backward scattering \citep{2025PhRvD.111f3055N,2025arXiv251012869N}. Compared with the unmagnetized case, the growth rate is suppressed by a factor of $(\omega_0/\omega_{\mathrm c})^{4}$ for ICS and by $(\omega_0/\omega_{\mathrm c})^{\frac{4}{3}}$ for SBS, which can be interpreted as a gyroradius effect \citep{2025PhRvD.111f3055N,2025arXiv251012869N}.

\section{PIC simulation setup}
We verify the linear growth rate of ICS driven by a circularly polarized Alfv\'en incident wave in magnetized $e^\pm$ plasma using the same PIC simulation scheme as in the companion paper \citet{2026arXiv260101169K}. The incident wave is initialized inside the periodic plasma domain.

We adopt plasma parameters that differ from those in \citep{2026arXiv260101169K} so that the dominant induced scattering process changes. The magnetization of the $e^\pm$ plasma is set to $\sigma_B\equiv B_0^2/(8\pi n_{\mathrm{e}0}m_{\mathrm{e}}c^2)=\omega_{\mathrm{c}}^2/\omega_{\mathrm{p}}^2=2$, the normalized incident frequency is $\omega_0/\omega_{\rm p}=7.1\times10^{-2}$, and the thermal velocity of electrons and positrons is $\sqrt{k_{\mathrm{B}}T_{\mathrm{e}}/(m_{\mathrm{e}}c^2)}=0.04$. The parameters are chosen to test induced scattering in a regime not explored in \citep{2026arXiv260101169K}. With this parameter set, we verify the neutral-mode transition from ICS to SBS and examine nonlinear evolution in which the incident wave is substantially attenuated, unlike the low-level saturation found in \citep{2026arXiv260101169K}. This nonlinear evolution corresponds to the high density (multiplicity), full scattering region in Fig.~\ref{Application_FRB_letter}.

The spatial resolution of the incident wavelength is $\lambda_0/\Delta x=1200$, and the simulation box size is $2\Delta x \times 32\lambda_0$ in the $(x\times y)$ directions, so that $L_y \gg L_x$ and the system effectively follows one-dimensional propagation along $\bm B_0$ (see Fig.~2 in \citep{2026arXiv260101169K}). This box size comfortably satisfies the resolution requirement for the scattered wavenumber, $L_y/\lambda_0\gtrsim 13$. The background magnetic field is set to $\bm B_0\parallel \hat{\bm y}$, and the incident field is a right-handed (electron-like) circularly polarized Alfv\'en wave with $\bm k_0\parallel \bm B_0$.

We initialize the particle velocity field, current, electric field, and magnetic field self-consistently from the finite-amplitude circularly polarized Alfv\'en-wave solution. This procedure avoids artificially seeding a backward-propagating scattered wave in the initial condition. The initial condition uses the exact cold plasma velocity field for a finite amplitude Alfv\'en wave. Since this velocity field is not an exact finite temperature solution, this mismatch produces a short bulk relaxation at the beginning of the run, but the associated change in the thermal velocity is negligible.

The one-dimensional setup was adopted because it allows us to follow the most important process in detail over a long period of time. The analytic theory shows that density fluctuations with $\bm{k}\perp\bm{B}_0$ do not lead to the growing induced-scattering modes considered here (see Appendix B in \citep{2025PhRvD.111f3055N}). This suggests that, even when oblique scattering occurs, the density fluctuations with $\bm{k}\parallel\bm{B}_0$ may play the central role. Direct PIC simulations of oblique scattering and multidimensional nonlinear evolution are left for future work.

\section{PIC simulation results}
\label{sec:PIC}
\begin{figure}
  \centering
  \includegraphics[width=\columnwidth]{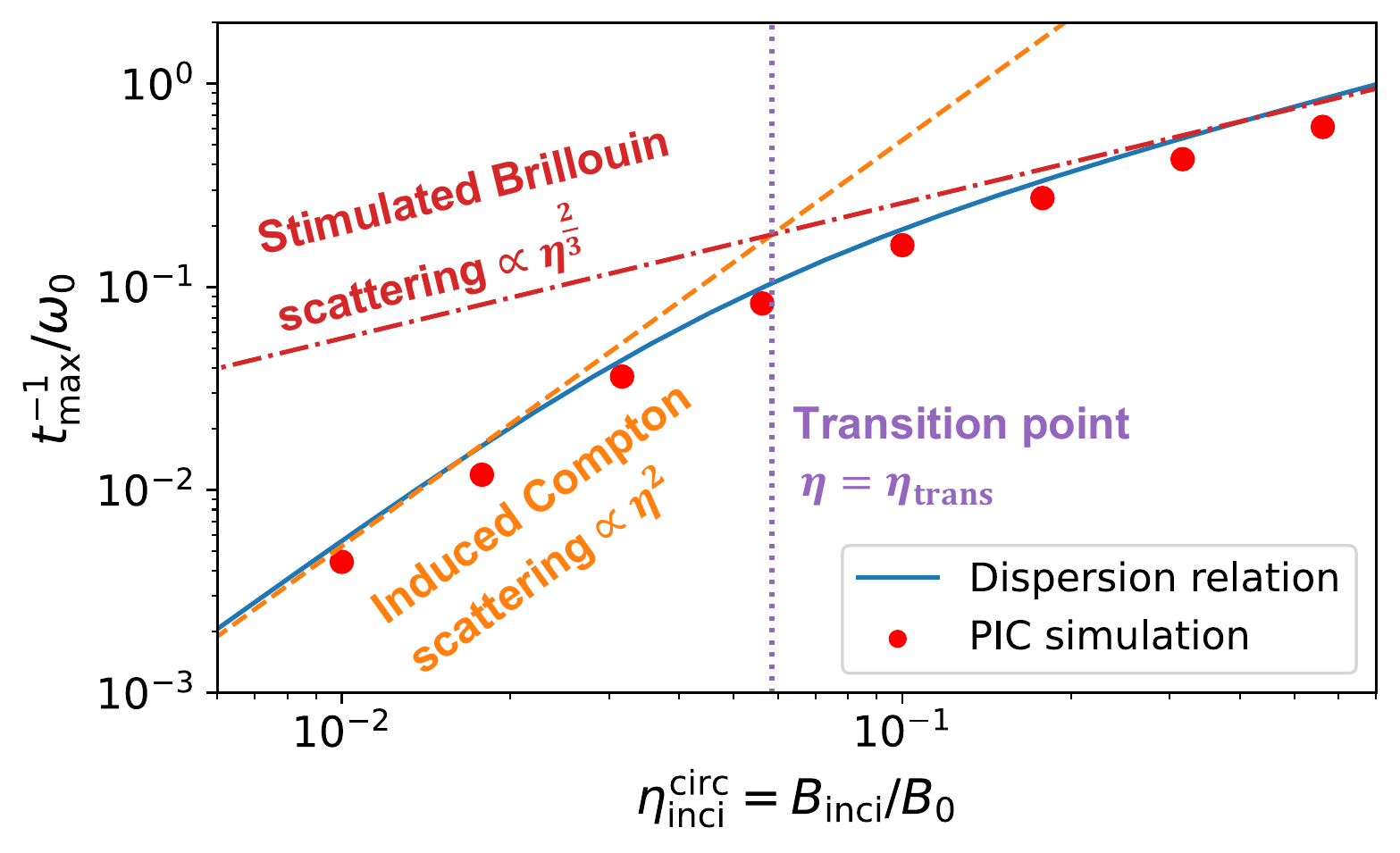}
  \caption{\justifying
  Maximum linear growth rate of induced scattering driven by the neutral mode for a monochromatic incident wave as a function of the incident wave amplitude $\eta^{\mathrm{circ}}_{\mathrm{inci}}$. Red dots denote the PIC results. The blue solid curve is the numerical solution of the dispersion relation of the neutral mode (see Eq.~(21) of \citep{2025arXiv251012869N}). The orange dashed and red dash-dotted curves are the analytic ICS and SBS in Eq.~\eqref{eq:growth_rate_neutral_matome_Brillouin_letter}, respectively. The purple dotted curve shows the ICS–SBS transition amplitude obtained from Eq.~\eqref{eq:transition_point_for_neutral_mode_letter} rewritten using Eq.~\eqref{eq:definition_of_eta_letter}.}
  \label{fig:PIC_neutral_growth_rate}
\end{figure}

Fig.~\ref{fig:PIC_neutral_growth_rate} shows that the maximum linear growth rate of the neutral mode agrees with the analytic expressions in Eq.~\eqref{eq:growth_rate_neutral_matome_Brillouin_letter}, with the solution of the dispersion relation (Eq.~(21) of \citep{2025arXiv251012869N}), and with the PIC measurements, and that it increases smoothly from ICS to SBS as $\eta^{\mathrm{circ}}_{\mathrm{inci}}$ grows. Although the SBS growth rate is larger than that of ICS on the small $\eta^{\mathrm{circ}}_{\mathrm{inci}}$ side, SBS is not realized because the acoustic wave in $e^\pm$ plasma is a heavily Landau-damped quasi-mode \citep{2025arXiv251012869N}. The deviations from the dispersion relation solution are only $16$--$28\%$ across all points.

\ioka{The agreement between the PIC results and the analytic formulae of the neutral mode in Eq. \eqref{eq:growth_rate_neutral_matome_Brillouin_letter} over a wide range of magnetization $\sigma_B$ also supports the predicted dependence on the background magnetic field $B_0$. In the present simulations, the maximum linear growth rate agrees with the analytic formulae for $\sigma_B=2$. In contrast, \citet{2026arXiv260101169K} found similar agreement for the much larger magnetization $\sigma_B=1250$. The consistency at these widely separated values of $\sigma_B$ therefore supports the dependence on $B_0$ predicted by the analytic theory.}

\begin{figure}
  \centering
  \includegraphics[width=\columnwidth]{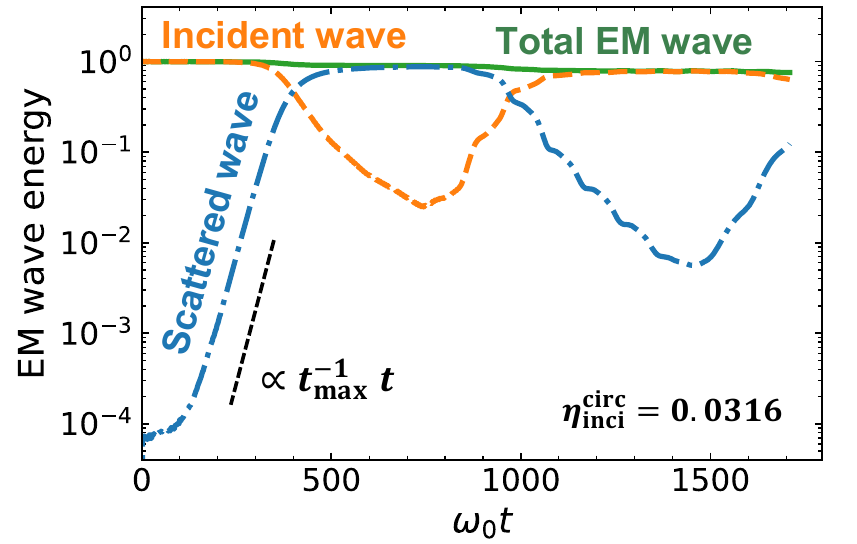}
  \caption{\justifying
  Time evolution of the forward and backward components of the EM wave energy, normalized by the initial incident wave energy. The horizontal axis is time normalized by $\omega_{0}$, and the vertical axis is the EM wave energy Fourier-transformed along $\bm{B}_0$ (see \citep{2026arXiv260101169K} for the detailed treatment of the Fourier transformation). The forward right-handed circularly polarized Alfv\'en waves (orange dashed) are initialized at $t=0$, and the scattered components (blue dot-dashed) are the backward right-handed circularly polarized Alfv\'en waves generated by induced scattering. The green solid curve shows the total EM wave energy, and the black dashed line indicates the slope corresponding to the maximum linear ICS growth rate of the neutral mode in Eq. \eqref{eq:growth_rate_neutral_matome_Brillouin_letter}.}
  \label{fig:energy_distribution_letter}
\end{figure}

The PIC simulation reveals the nonlinear evolution beyond the linear growth stage. For the small values of $\eta^{\mathrm{circ}}_{\mathrm{inci}}$ in Fig.~\ref{fig:PIC_neutral_growth_rate}, corresponding to the leftmost five points, the forward and backward components undergo oscillatory energy exchange, as shown in Fig.~\ref{fig:energy_distribution_letter}. \ioka{Appendix~\ref{app:spatially_resolved} also presents the spatially resolved evolution of the electric field components of the EM waves.} At early times, the backward waves grow in accordance with the linear growth rate in Eq.~\eqref{eq:growth_rate_neutral_matome_Brillouin_letter}\ioka{, as also seen in the spatial distribution in Figs.~\ref{fig:spatial_entire}(b) and \ref{fig:spatial_figure}(a)}. The total EM wave energy decreases because it is transferred to the internal energy of the plasma, but the decrease is small. The backward waves then grow until they reach an energy comparable to that of the forward wave, while the forward wave is strongly attenuated\ioka{, as illustrated in Figs.~\ref{fig:spatial_entire}(c) and \ref{fig:spatial_figure}(b)}. The time required for the backward wave energy to first become comparable to the forward wave energy is approximately ten e-foldings. This estimate holds for all of the first five data points from the left in Fig.~\ref{fig:PIC_neutral_growth_rate}, without dependence on $\eta_{\mathrm{inci}}$. For example, for $\eta_{\mathrm{inci}}=0.0316$, the third data point from the left in Fig.~\ref{fig:PIC_neutral_growth_rate} gives the linear growth rate as $t_{\mathrm{max}}^{-1}/\omega_0\sim0.036$. Fig.~\ref{fig:energy_distribution_letter} shows that the forward and backward wave energies first become comparable at $\omega_0t_{\mathrm{sat}}\sim400$. These values give $t_{\mathrm{max}}^{-1}t_{\mathrm{sat}}\sim14$, which is consistent with saturation after approximately ten e-foldings. Subsequently, the roles of the two components are exchanged. The forward waves (granddaughter waves) increase again, whereas the backward waves decrease\ioka{, as shown in Figs.~\ref{fig:spatial_entire}(d), \ref{fig:spatial_figure}(c), and (d)}. Note that the frequency and wavenumber of the granddaughter waves differ from those of the incident wave. \ioka{At still later times, the backward waves begin to grow again, indicating that the nonlinear energy exchange continues beyond the first conversion cycle, as shown in Fig.~\ref{fig:spatial_figure}(e).} Although the total EM wave energy appears to be nearly conserved within a local simulation box, an FRB propagates through fresh scattering regions, so the local energy exchange cannot complete an oscillation cycle in each region. On average, the FRB is strongly damped, and we refer to this case as \emph{full scattering}.

In contrast, the three largest values of \(\eta^{\mathrm{circ}}_{\mathrm{inci}}\) in Fig.~\ref{fig:PIC_neutral_growth_rate} show a qualitatively different evolution. The scattered wave initially grows linearly, but then saturates after roughly ten e-foldings, and no oscillatory energy exchange is observed. We refer to this behavior as \emph{partial scattering}. A similar low-level saturation is seen in Fig.~5 of \citep{2026arXiv260101169K} for different parameters.

Induced scattering saturates when a plateau forms in the plasma distribution function around the phase velocity of the beat wave \citep{2026arXiv260101169K}. The distribution function evolves according to the Vlasov equation \eqref{eq:Vlasov_eq_letter}, in which the ponderomotive force enters as $-(1/m_{\mathrm{e}})\,\bm{\nabla}\phi_{\mathrm{p}}^{\pm}\cdot(\partial f_{\pm}/\partial\bm{v})$. As the instability grows, it flattens the slope of the distribution function near the phase velocity of the beat wave. As the slope becomes shallower and eventually approaches a plateau, the ponderomotive force can no longer sustain the instability, and the system saturates. This is essentially for the same reason that Landau damping saturates once the distribution function flattens into a plateau \citep{1969npt..book.....S,2017RvMPP...1....5M}.

A plateau is expected to form once the EM wave deposits an energy density comparable to the internal energy density of the plasma. The internal energy density is estimated as 
\begin{equation}
   \varepsilon_{\mathrm{th}} = 2 n_{\mathrm{e}0}\cdot \frac{1}{2} k_{\mathrm{B}} T_{\mathrm{e}}, 
\end{equation}
since only the velocity along $\bm{B}_{0}$ enters the resonance and the effective degree of freedom is $1$. In a single scattering event, energy conservation implies that an incident wave with energy $\omega_0$ transfers an energy fraction $\omega_1$ to the scattered wave and a fraction $|\omega_1-\omega_0|$ to the plasma. Then the energy density of the scattered wave at saturation is estimated as
\begin{equation}
    \varepsilon_{\mathrm{scat}}^{\mathrm{max}}\sim\frac{\omega_1}{|\omega_1-\omega_0|}\varepsilon_{\mathrm{th}}\,.
    \label{eq:saturation_condition_letter}
\end{equation}
For ICS, Eq.~(30) of \citep{2026arXiv260101169K} yields 
\begin{equation}
    \varepsilon_{\mathrm{scat}}^{\mathrm{max}}\sim\frac{1}{2}\frac{v_{\mathrm{A}}}{c}\left(\frac{m_{\mathrm{e}}c^2}{k_{\mathrm{B}}T_{\mathrm{e}}}\right)^{\frac{1}{2}}\varepsilon_{\mathrm{th}},
    \label{eq:saturation_condition_ICS}
\end{equation}
where 
\begin{equation}
\begin{aligned}
    v_{\mathrm{A}}&\equiv c\bigl(1+\omega_{\mathrm{p}}^{2}/\omega_{\mathrm{c}}^{2}\bigr)^{-\frac{1}{2}}\\
    &\sim0.82c\quad\mathrm{for~}\sigma_B=2,    
\end{aligned}
\end{equation}
is the Alfv\'en speed. For SBS, the angular frequency of the density fluctuation $\omega_1-\omega_0$ is derived in Eq.~(C6) of \citep{2025arXiv251012869N}. Taking its real part and using $\omega_1\sim\omega_0$, one obtains
\begin{equation}
\varepsilon_{\mathrm{scat}}^{\mathrm{max}}\sim2\sigma_{B}^{\frac{1}{3}}\eta_{\mathrm{inci}}^{-\frac{2}{3}}\varepsilon_{\mathrm{th}}.
\label{eq:saturation_SBS}
\end{equation}

The saturation energy density estimated in Eq.~\eqref{eq:saturation_condition_letter} agrees with the PIC simulation results within a factor of three. We define the energy density of the incident wave as
\begin{equation}
\varepsilon_{\mathrm{inci}}
\equiv
\frac{|\bm{E}_{\mathrm{inci}}|^{2}+ |\bm{B}_{\mathrm{inci}}|^{2}}{8\pi}.
\end{equation}
For the parameters used in Fig.~\ref{fig:energy_distribution_letter}, the ratio of the incident wave energy density to the saturation energy density of the scattered waves is expressed from Eq. \eqref{eq:saturation_condition_ICS} as
\begin{equation}
\begin{aligned}
\frac{\varepsilon_{\mathrm{inci}}}{\varepsilon_{\mathrm{scat}}^{\mathrm{max}}}
&\simeq0.24~
\left(\frac{\eta_{\mathrm{inci}}}{0.0316}\right)^{2}
\left(\frac{\sigma_B}{2}\right)
\\
&\quad \times
\left(\frac{\sqrt{k_BT_{\mathrm{e}}/(m_\mathrm{e}c^2)}}{0.04}\right)^{-1}
<1.
\end{aligned}
\label{eq:saturation_check}
\end{equation}
This inequality implies that the incident wave energy is insufficient to reach the total scattered wave energy expected at saturation. The incident wave is therefore converted into scattered waves before nonlinear saturation, leading to the oscillatory energy exchange shown in Fig.~\ref{fig:energy_distribution_letter}. For the SBS case with the amplitude $\eta_{\mathrm{inci}}=0.562$, corresponding to the rightmost data point in Fig.~\ref{fig:PIC_neutral_growth_rate}, Eq.~\eqref{eq:saturation_SBS} gives
\begin{equation}
\begin{aligned}
\frac{\varepsilon_{\mathrm{inci}}}{\varepsilon_{\mathrm{scat}}^{\mathrm{max}}}
&\simeq
2.1\times10^2~
\left(\frac{\eta_{\mathrm{inci}}}{0.562}\right)^{\frac{8}{3}}
\left(\frac{\sigma_B}{2}\right)^{\frac{2}{3}}
\\
&\quad \times
\left(\frac{\sqrt{k_BT_{\mathrm{e}}/(m_\mathrm{e}c^2)}}{0.04}\right)^{-2}
\gg1.
\end{aligned}
\label{eq:saturation_check_SBS}
\end{equation}
The incident wave therefore has more energy than the total scattered wave energy required for saturation. In this regime, the instability reaches saturation with only weak attenuation of the incident wave. Indeed, in the PIC simulation, the total scattered wave energy saturates at $\varepsilon_{\mathrm{scat}}^{\mathrm{sim}}/\varepsilon_{\mathrm{scat}}^{\mathrm{max}}\sim2.6$, which agrees with Eq.~\eqref{eq:saturation_SBS} within a factor of three. A more detailed analysis, including the time evolution of the plasma distribution function, will be presented in future work.

%\ioka{We find that the ICS saturation estimate in Eq.~\eqref{eq:saturation_condition_ICS} agrees with the total scattered-wave energy at saturation within a factor of two. To check the saturation estimate, we reanalyzed a representative neutral-mode ICS run, Run 16 of \citet{2026arXiv260101169K}, which exhibits nonlinear saturation. The right panel of Fig.~5 in \citep{2026arXiv260101169K} shows the time evolution of the scattered-wave energy in the single Fourier component with the maximum linear growth rate. 
%If we compare this single-component energy with Eq.~\eqref{eq:saturation_condition_ICS}, the PIC result is smaller by about an order of magnitude. When the energy is summed over all growing scattered components, the total scattered-wave energy agrees with Eq.~\eqref{eq:saturation_condition_ICS} within a factor of two \footnote{\ioka{The right panel of Fig.~4 in \citet{2026arXiv260101169K} shows that the scattered waves grow over a finite range of wavenumbers around the maximum growth wavenumber, indicated by the dashed cyan line, rather than only in a single Fourier component.}}. 
%Thus, Eq.~\eqref{eq:saturation_condition_letter} should be interpreted as an order-of-magnitude estimate for the total scattered wave energy at saturation, rather than for a single Fourier component.}

\section{Linear growth rates for a broadband incident wave}
In the PIC simulations, we adopted a monochromatic incident wave. When applying our results to FRBs, however, it is necessary to account for the finite bandwidth $\Delta\omega$ of the incident wave. Some FRBs exhibit broadband emission with $\Delta\omega\sim\omega_0$ \citep{2021ApJ...923....1P}, and we therefore consider broadband incident waves. Finite bandwidth reduces phase coherence between the incident and scattered waves and weakens the effective three wave resonance. This reduction lowers the maximum linear growth rate relative to the monochromatic limit \citep{1974PhFl...17..849T,1994ApJ...422..304T,kruer2019physics,2021PPCF...63i4003B,2023RvMPP...7....1Z}. For the neutral mode, the growth rate scales as $(t_{\mathrm{neutral}}^{\mathrm{broad}})^{-1}\propto a_{\mathrm e}^{2}\Delta\omega^{-2}$ in the ICS regime and as $(t_{\mathrm{neutral}}^{\mathrm{broad}})^{-1}\propto a_{\mathrm e}^{\frac{4}{3}}\Delta\omega^{-1}$ in the SBS regime. The explicit growth rate formulae and the ICS–SBS transition amplitude are summarized in Eqs.~\eqref{eq:neutral_summary_letter} and \eqref{eq:weak_strong_transition_inc_neutral_letter}.

For the charged mode, the behavior of induced scattering depends strongly on the plasma density \citep{2025arXiv251012869N}. In the regime $\omega_0 \ll \omega_{\mathrm p} \ll \omega_{\mathrm c}$, ICS with Debye screening is the dominant process. The corresponding maximum linear growth rate for a broadband incident wave is listed in Eq.~\eqref{eq:Compton_charged_Brillouin_growth_rate_Debye_letter}, and scales as $(t_{\mathrm{charged}}^{\mathrm{broad}})^{-1}\propto a_{\mathrm e}^{2}\Delta\omega^{-2}$. Compared with the unmagnetized case, the linear growth rate is suppressed not only by the gyroradius effect $(\omega_0/\omega_{\mathrm{c}})^2$ but also by the Debye screening effect $(\omega_0/\omega_{\mathrm{p}})^4(8k_{\mathrm{B}}T_{\mathrm{e}}/(m_{\mathrm{e}}c^2))^2(1+\omega_{\mathrm{p}}^2/\omega_{\mathrm{c}}^2)^2$ \citep{2025PhRvD.111f3055N}.

\section{Application to extragalactic FRBs}
\label{sec:Application}
\begin{figure}
  \centering
  \includegraphics[width=\columnwidth]{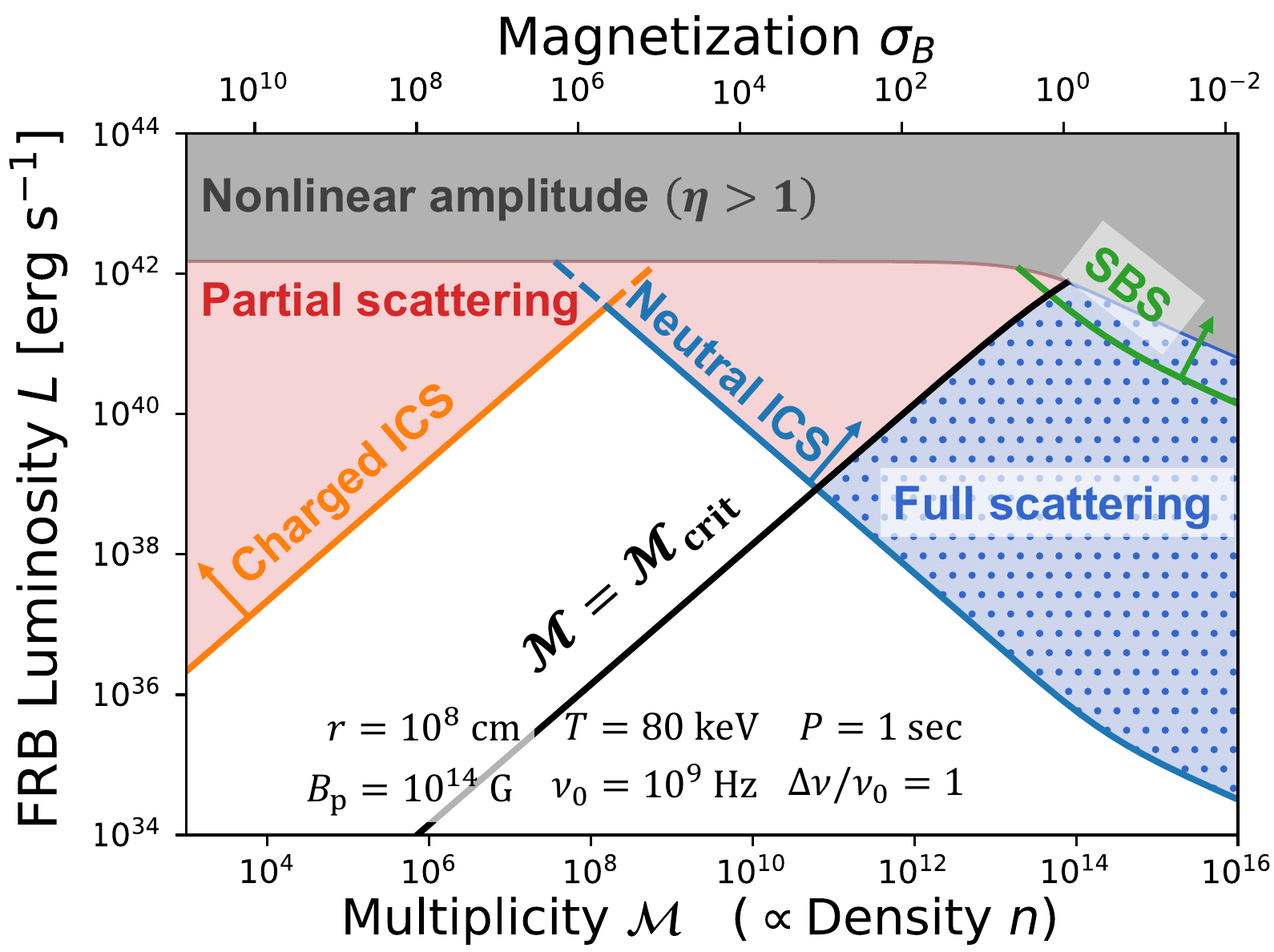}
  \caption{\justifying Regime map of induced scattering for GHz FRB pulses propagating through a dipolar magnetar magnetosphere. The horizontal axis shows the multiplicity $\mathcal{M}$, and the top axis shows the corresponding magnetization parameter $\sigma_B$. The vertical axis shows the FRB Poynting luminosity $L$. Colored/shaded regions indicate partial scattering (red), full scattering (blue hatched), and the nonlinear amplitude $\eta>1$ (gray). The white region denotes parameters for which neither the charged nor the neutral mode reaches the linear growth stage within a FRB duration $\Delta t$. The dominant process in each region is labeled as “Neutral ICS,” “Charged ICS,” or “(Neutral) SBS”.\footnote{\justifying At high $\mathcal{M}$ ($\sigma_B\lesssim 1$), the curves bend because the subluminal correction $(1+\omega_{\mathrm{p}}^2/\omega_{\mathrm{c}}^2)=(c/v_{\mathrm{A}})^2$ becomes non-negligible.} The black curve marks the boundary between partial and full scattering. The parameters used in this figure are listed in the panel.} 
  \label{Application_FRB_letter}
\end{figure}
\begin{figure}
  \centering
  \includegraphics[width=\columnwidth]{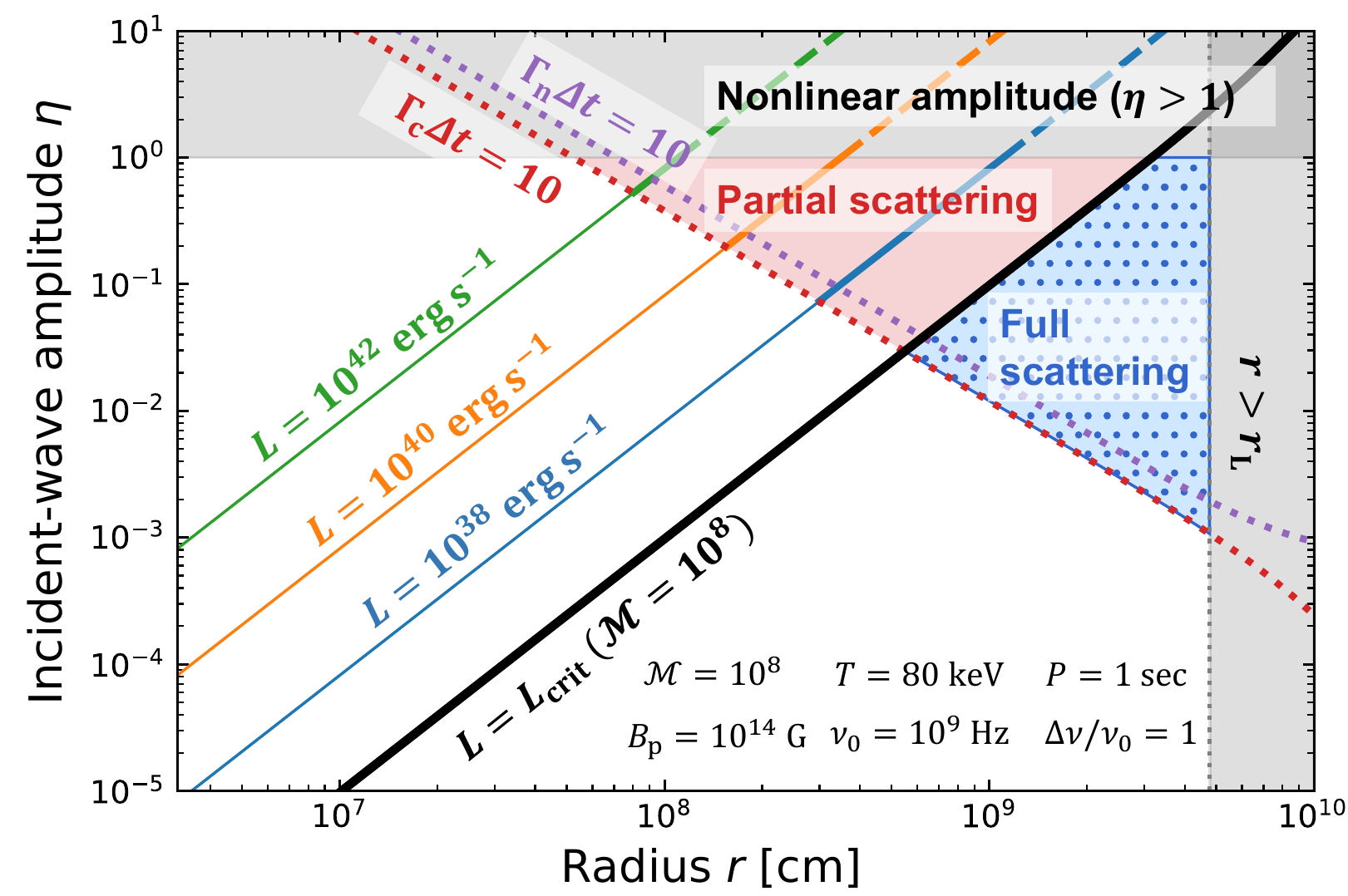}
  \caption{\justifying Radial dependence of induced scattering for FRBs propagating through a dipolar magnetar magnetosphere. The solid colored curves show the incident-wave amplitude $\eta(r)$ for $L=10^{38}$, $10^{40}$, and $10^{42}\,{\rm erg\,s^{-1}}$. Thin parts indicate radii before the wave enters the linear growth stage, while thick parts indicate radii where at least one induced-scattering mode satisfies $\Gamma\Delta t>10$, where $\Gamma\equiv (t^{\rm broad})^{-1}$. The dotted red and purple curves show the charged and neutral linear-growth thresholds, $\Gamma_{\rm c}\Delta t=10$ and $\Gamma_{\rm n}\Delta t=10$, respectively. The black curve denotes the boundary between partial and full scattering. Red and blue hatched regions show the partial and full scattering regimes. The white region denotes parameters for which neither the charged nor the neutral mode reaches the linear growth stage within a FRB duration $\Delta t$. Gray regions indicate either the nonlinear amplitude, $\eta>1$ or radii outside the light cylinder, $r>r_{\rm L}\equiv c/\Omega$.}
  \label{radial_dependence}
\end{figure}

We now apply the above framework to GHz FRBs propagating as fast magnetosonic waves in a dipolar magnetar magnetosphere and summarize the main result in Figs.~\ref{Application_FRB_letter} and \ref{radial_dependence}. For a dipole field, the magnetic field strength at radius $r$ can be expressed in terms of the polar surface field $B_{\mathrm p}$ as $B(r)\sim B_{\mathrm{p}}R^{3}/r^{3}\sim10^{8}~\mathrm{G}~r_8^{-3}R_6^{3}B_{\mathrm{p},14}$, and the plasma density is given by the Goldreich--Julian density \citep{1969ApJ...157..869G} multiplied by the pair multiplicity $\mathcal{M}$,  
$n(r)\equiv\mathcal{M}n_{\mathrm{GJ}}(r)\sim \mathcal{M}B(r)/(ceP)\sim6.9\times10^{14}~\mathrm{cm}^{-3}~P^{-1}r_8^{-3}R_6^{3}B_{\mathrm{p},14}\mathcal{M}_8$. Here $R$ and $P$ are the magnetar radius and rotation period, and we use the notation $A\equiv10^{n}A_{n}$ in CGS units and $T_{\mathrm{80keV}}\equiv T/\mathrm{80keV}$ for temperatures. This temperature is close to the cutoff energy inferred for the X-ray burst associated with the Galactic FRB \citep{2021NatAs...5..378L}.

We apply the growth rates derived for a uniform $\bm{B}_0$, as local rates in a dipolar magnetosphere. This local treatment is justified to order of magnitude when $\bm{B}_0$ variation scale is larger than the characteristic length scales of the FRB. There are two relevant length scales. The first is the wave scale. For a dipole field, the magnetic-field scale height is $H_B\equiv|\dd\ln B_0/\dd r|^{-1}=r/3\sim3\times10^7~\mathrm{cm}~r_8$, which is much larger than the wavelength of a GHz wave, $\lambda_0\sim30\,{\rm cm}~\nu_9^{-1}$. The second is the pulse scale. For a millisecond pulse, $c\Delta t\sim3\times10^7\,{\rm cm}$, which is comparable to $H_B$ at $r\sim10^8\,{\rm cm}$. Although $\bm{B}_0$ can vary across the pulse at smaller radii, induced scattering is strongly suppressed there by the larger $\bm{B}_0$, so this radial variation does not substantially change our order-of-magnitude estimate.

Once $B(r)$ and $n(r)$ are specified, the cyclotron and plasma frequencies are determined. We consider an FRB with central frequency $\nu_0=\omega_0/(2\pi)$, bandwidth $\Delta\nu$, duration $\Delta t$, and Poynting luminosity $L\equiv4\pi r^{2}v_{\mathrm{A}}\langle|\bm{E}_{\mathrm{w}}|^{2}+|\bm{B}_{\mathrm{w}}|^{2}\rangle/(8\pi)$. The strength parameter is then given by  
$a_{\mathrm e}=2eL^{\frac{1}{2}}(1+c^{2}/v_{\mathrm{A}}^{2})^{-\frac{1}{2}}/(m_{\mathrm{e}}cv_{\mathrm{A}}^{\frac{1}{2}}\omega_{0}r)\simeq2.3\times10^{4}\,r_8^{-1}L_{40}^{\frac{1}{2}}\nu_9^{-1}$. Since $\omega_{\mathrm p}/\omega_{\mathrm c}=1/\sqrt{\sigma_B}\sim1.2\times10^{-3}\,\mathcal{M}_{8}^{\frac{1}{2}}\,r_8^{\frac{3}{2}}P^{-\frac{1}{2}}R_6^{-\frac{3}{2}}B_{\mathrm{p},14}^{-\frac{1}{2}}$, the subluminal effect $(1+\omega_{\mathrm{p}}^{2}/\omega_{\mathrm{c}}^{2})=(c/v_{\mathrm{A}})^{2}\sim1$ can be neglected. Once $a_{\mathrm e}$ is specified, the dimensionless amplitude of the FRB is obtained as $\eta\simeq a_{\mathrm e}\omega_{0}/\omega_{\mathrm c}\simeq 8.2\times10^{-2}\,r_8^{2}L_{40}^{\frac{1}{2}}B_{\mathrm{p},14}^{-1}R_6^{-3}$.

Note that the region below the green SBS curve in Fig.~\ref{Application_FRB_letter} can be treated only with the kinetic treatment developed in this work rather than fluid treatment. The condition for the dominant induced scattering in the neutral mode to be ICS rather than fluid-type SBS is from Eq. \eqref{eq:weak_strong_transition_inc_neutral_letter}, $\mathcal{M}<2.3\times10^{15}\,PR_6^{9}B_{\mathrm{p},14}^{3}r_8^{-7}L_{40}^{-1}(\Delta\nu/\nu_0)^3$. As shown in Fig.~\ref{fig:PIC_neutral_growth_rate}, extrapolating the SBS growth rate into this ICS-dominated region would overestimate the growth rate by several orders of magnitude.

The maximum linear growth rates of the charged and neutral modes for a broadband FRB (Eqs. \eqref{eq:Compton_charged_Brillouin_growth_rate_Debye_letter} and the ICS regime of~\eqref{eq:neutral_summary_letter}) are estimated as
\begin{equation}
\begin{aligned}
\left(t_{\mathrm{charged}}^{\mathrm{broad}}\right)^{-1}
&\approx 4.6\times10^{2}~\mathrm{s^{-1}}\,\frac{P r_{8}^{7} L_{40}T_{\mathrm{80keV}}^2\nu_{9}^{3}}{\mathcal{M}_{8}  R_{6}^{9} B_{\mathrm{p},14}^{3}}
\left( \frac{ \Delta \nu}{\nu_0} \right)^{-2},
\\
\left(t_{\mathrm{neutral}}^{\mathrm{broad}}\right)^{-1}
&\approx 1.9\times10^{2}~\mathrm{s^{-1}}\,\frac{\mathcal{M}_{8} r_{8}^{7}L_{40} \nu_{9}}{P  R_{6}^{9}B_{\mathrm{p},14}^{3}}
\left( \frac{ \Delta \nu}{\nu_0} \right)^{-2},
\end{aligned}
\label{eq:induced_scattering_FRB_letter}
\end{equation}
and we compare these with a typical FRB inverse duration $\Delta t^{-1}\sim10^{3}\,\mathrm{s^{-1}}$. The red and blue hatched regions in Fig.~\ref{Application_FRB_letter} show where at least one mode of induced scattering grows by more than ten e-foldings at $r=10^8~\mathrm{cm}$. This condition is written as $(t^{\mathrm{broad}})^{-1} > 10\Delta t^{-1}$.
The numerical factor of 10 is motivated by the PIC simulations. As discussed in Sec.~\ref{sec:PIC}, the simulations show that approximately ten e-foldings are required both for induced scattering to saturate in the partial scattering and for the incident wave to be attenuated in the full scattering.

At first glance, most FRBs would appear to be strongly scattered. However, our results indicate a wide parameter region where induced scattering saturates, so that FRBs undergo only partial scattering. This occurs when the total FRB energy exceeds the saturation energy of the scattered wave in Eq. \eqref{eq:saturation_condition_letter}. The total energy of an FRB is written as $\mathcal{E}_{\mathrm{FRB}}=L\Delta t\sim10^{37}~\mathrm{erg}~L_{40}\Delta t_{-3}$. The saturation energy of the scattered wave at radius $r$ is obtained from Eq.~\eqref{eq:saturation_condition_letter} and is estimated as $\mathcal{E}_{\mathrm{scat}}^{\mathrm{max}}\sim 4\pi r^2\Delta r\,\varepsilon_{\mathrm{scat}}^{\mathrm{max}}\sim 4\pi r^3\,\varepsilon_{\mathrm{scat}}^{\mathrm{max}}$. Therefore, the condition $\mathcal{E}_{\mathrm{FRB}}>\mathcal{E}_{\mathrm{scat}}^{\mathrm{max}}$ implies partial scattering, while $\mathcal{E}_{\mathrm{FRB}}\leq\mathcal{E}_{\mathrm{scat}}^{\mathrm{max}}$ implies full scattering. At the boundary $\mathcal{E}_{\mathrm{FRB}}=\mathcal{E}_{\mathrm{scat}}^{\mathrm{max}}$, this criterion can be written, for ICS, as a critical luminosity, 
\begin{equation}
L_{\mathrm{crit}}=\frac{2\pi r^3 n(r) m_{\mathrm e}c^2}{\Delta t}
    \sqrt{\frac{k_{\mathrm B}T_{\mathrm e}}{m_{\mathrm e}c^2}}.    
\end{equation}
Solving for $n(r)$ yields the critical multiplicity shown by the black solid curve in Fig.~\ref{Application_FRB_letter}, and it is calculated as
\begin{equation}
\begin{aligned}
    \mathcal{M}_{\mathrm{crit}}
    &\equiv\frac{1}{6}\frac{L\Delta t}{\frac{4\pi r^3}{3}\frac{B(r)^2}{8\pi}}
    \frac{\omega_{\mathrm{c}}(r)}{\Omega}
    \sqrt{\frac{m_{\mathrm{e}}c^2}{k_{\mathrm{B}}T_{\mathrm{e}}}}\\
    &\sim 7.1\times10^{11}~\frac{L_{40} P \Delta t_{-3}}{R_6^3 B_{\mathrm{p}, 14} T_{80 \mathrm{keV}}^{\frac{1}{2}}}\,.
\end{aligned}
\label{eq:critical_multiplicity}
\end{equation}
The large $\mathcal{M}_{\mathrm{crit}}$ arises because the ratio $\omega_{\mathrm{c}}(r)/\Omega$ is large, where $\Omega\equiv2\pi/P$ is the angular velocity of the magnetar. Although this density is naively Thomson thick, Thomson scattering for GHz EM waves with $\bm{E}_{\mathrm{w}0}\perp\bm{B}_0$ is suppressed by $(\omega_0/\omega_{\mathrm{c}})^2$ \citep{2024PhRvD.109d3048N}.

Fig.~\ref{radial_dependence} shows that induced scattering becomes important at larger radii for typical FRB luminosities before the wave reaches the nonlinear amplitude $\eta=1$. The green, orange, and blue curves show the radial evolution of the incident wave amplitude $\eta$ for $L=10^{38}$, $10^{40}$, and $10^{42}~\mathrm{erg~s}^{-1}$, respectively. The amplitude $\eta$ increases outward as $\eta\propto r^2$, even though the usual strength parameter decreases as $a_{\mathrm{e}}\propto r^{-1}$. This is because the background magnetic field decreases more rapidly, $B(r)\propto r^{-3}$. As a result, FRBs with typical luminosity $L\sim10^{38-42}~\text{erg~s}^{-1}$ always enters the linear growth stage of induced scattering before the wave reaches $\eta=1$. Thus, the nonlinear evolution of induced scattering must be considered before discussing the nonlinear amplitude regime, $\eta>1$.

Whether the nonlinear evolution is partial or full scattering is controlled mainly by the plasma density. As shown by Eq.~\eqref{eq:critical_multiplicity}, the boundary between the two regimes depends on the multiplicity $\mathcal{M}$. Fig.~\ref{radial_dependence} shows the case $\mathcal{M}=10^8$. For a given $\mathcal{M}$, an FRB with $L>L_{\mathrm{crit}}$ undergoes partial scattering, whereas one with $L<L_{\mathrm{crit}}$ undergoes full scattering after the linear growth stage. Increasing $\mathcal{M}$ raises $L_{\mathrm{crit}}$ and therefore makes full scattering more likely at fixed luminosity. The multiplicity $\mathcal{M}$ is one of the most uncertain parameters in magnetar magnetosphere.
The plasma density can differ greatly between a quiescent magnetosphere and a burst or fireball environment. Thus, low plasma density, as expected in a relatively quiescent magnetosphere, favors partial scattering, whereas strong plasma loading during a burst favors full scattering.

Changing the spin period mainly shifts the boundary between partial and full scattering, while the radii at which induced scattering enters the linear growth stage are almost unchanged. This dependence is shown in Figs.~\ref{fig:LMmap_sec} and \ref{fig:radial_sec} in Appendix~\ref{app:spin_period}, where we present the local $L$--$\mathcal{M}$ regime maps and radial dependence maps for $P=0.1\,{\rm s}$ and $P=10\,{\rm s}$. The weak dependence of the linear growth radii follows from Eq.~\eqref{eq:induced_scattering_FRB_letter}, since these radii depend on $P$ only through $P^{\pm \frac{1}{7}}$. In contrast, the boundary between partial and full scattering scales approximately as $\mathcal{M}_{\rm crit}\propto P$ from Eq. \eqref{eq:critical_multiplicity}. Thus, full scattering becomes more likely for a more rapidly rotating magnetar.

\section{Discussions}

Our results have important observational implications. Whether induced scattering strongly attenuates FRBs in magnetar magnetospheres depends on the plasma density. For typical FRB luminosities $L\sim10^{38-42}~\text{erg/s}$, the waves generally enter the linear growth stage of ICS irrespective of density, but if the plasma density is not extremely dense, $\mathcal{M}<\mathcal{M}_{\mathrm{crit}}=7.1\times10^{11}~L_{40} P \Delta t_{-3}R_6^{-3} B_{\mathrm{p}, 14}^{-1} T_{80 \mathrm{keV}}^{-\frac{1}{2}}$, induced scattering can saturate before causing substantial attenuation. The nonlinear evolution of induced scattering must therefore be taken into account before considering the nonlinear amplitude regime, $\eta>1$. This is consistent with observational indications that some extragalactic FRBs have compact source sizes comparable to magnetar magnetospheres \citep{2022Natur.607..256C,2022NatAs...6..393N,2025Natur.637...48N}. In contrast, for magnetar giant flares, observations suggest that a large amount of matter is ejected \citep{2005Natur.434.1104G,2005ApJ...634L..89G,2006ApJ...638..391G}, implying a high-density region $\sigma_B\sim1$ where full scattering is consistent with the non-detection of associated FRBs. A similar argument applies when a fireball forms during magnetar bursts \citep{1995MNRAS.275..255T,2020ApJ...904L..15I,2023MNRAS.519.4094W}. The plasma density can become high, $\mathcal{M}>\mathcal{M}_{\mathrm{crit}}$, so the system enters the full scattering regime and the FRB is strongly damped. This may explain why most X-ray bursts lack accompanying FRBs \citep{2021ApJ...906L..12Y,2021NatAs...5..414K}.

%\ioka{The relation to previous escape arguments should be interpreted with care. In the subnonlinear-amplitude, $\eta<1$, \citet{2025PhRvD.111f3055N} derived the linear growth rates of ICS in strongly magnetized $e^\pm$ plasma, and \citet{2025arXiv251012869N} analyzed the competition among ICS, SBS, and SRS. The present work applies these growth rates to FRBs propagating through a magnetar magnetosphere and follows the nonlinear evolution.}

Our results show that we must consider the nonlinear evolution of induced scattering before discussing nonlinear FRB amplitudes with $\eta>1$. The present study provides the basis for $\eta>1$. Even in linear amplitude, $\eta<1$, future work should examine competition with other parametric instabilities, including interactions between Alfv\'en and fast magnetosonic waves \citep{1998PhRvD..57.3219T,2019ApJ...881...13L,2019MNRAS.483.1731L,2023ApJ...957..102G}. It is then necessary to explore how induced scattering evolves after the wave amplitude reaches $\eta>1$. Previous studies suggested that a fast magnetosonic wave with $\eta\sim1$ can be damped by nonlinear shock formation \citep{2021ApJ...922L...7B,2022PhRvL.128y5003B,2023ApJ...959...34B,2024ApJ...975..223B}. However, it is also argued that shock formation no longer occurs as the propagation direction becomes closer to the background magnetic field \citep{2023ApJ...959...34B}. The behavior of a superluminal EM wave in the regime $\omega_{\mathrm{c}}/a_{\mathrm{e}}<\omega_{0}<a_{\mathrm{e}}\omega_{\mathrm{c}}$ also remains to be clarified \citep{2024A&A...690A.332S}. Applying the present framework to the propagation of pulsar radio emission \citep{1975Ap&SS..36..303B,1978MNRAS.185..297W,1982MNRAS.200..881W} and to precursor emission from neutron star mergers \citep{1996A&A...312..937L,2000ApJ...537..327I,2001MNRAS.322..695H,2020PTEP.2020j3E01W} is another important direction.

\begin{acknowledgments}
We gratefully acknowledge insightful discussions with Masanori Iwamoto, Jonathan Granot, Wataru Ishizaki, Takashi Hosokawa. RN is supported by JST SPRING, Grant No. JPMJSP2110, and JSPS KAKENHI, Grant No. 25KJ1562. KI is supported by MEXT/JSPS KAKENHI Grant No. 26H02045, 23H01172, 23H05430, 23H04900, 22H00130. SK is supported by MEXT/JSPS KAKENHI Grant No.22H00130 and 23K20038. \niShiura{Numerical computations were carried out on Cray XC50 and XD2000 at Center for Computational Astrophysics, National Astronomical Observatory of Japan, Yukawa-21 at YITP in Kyoto University, and Flow in Nagoya University through the HPCI System Research Project (Project ID: hp240147, hp250036)} The authors thank the Yukawa Institute for Theoretical Physics at Kyoto University, where this work was further developed during the YITP-W-25-08 on "Exploring Extreme Transients: Frontiers in the Early Universe and Time-Domain Astronomy".
\end{acknowledgments}
%\selectlanguage{english}
% --- Appendix ---
\clearpage
\appendix

\section{Linear Growth Rates}
\label{app:growth_rates}
The formulae summarized in this section were derived in \citep{2025PhRvD.111f3055N,2025arXiv251012869N}. We reproduce them here because they are directly used in the PIC comparison and in the FRB application.

\subsection{Neutral Mode: Linear Growth Rates for a Monochromatic Incident Wave}

The maximum linear growth rate of the neutral mode for a monochromatic incident wave is (see Eq. (66) in \citep{2025arXiv251012869N})
\begin{equation}
\left(t_{\text{neutral}}^{\text{coh}}\right)^{-1} \sim 
\begin{cases}
\sqrt{\frac{\pi}{32 \text{e}}} \frac{a_{\mathrm{e}}^{2} \omega_{\mathrm{p}}^{2}}{\omega_{0}} \frac{m_{\mathrm{e}} c^{2}}{k_{\mathrm{B}} T_{\mathrm{e}}}
\left(\frac{\omega_{0}}{\omega_{\mathrm{c}}}\right)^{4}
\left(1+\frac{\omega_{\text{p}}^2}{\omega_{\text{c}}^2}\right)^{-1}, 
\\
\quad a_{\mathrm{e}} \frac{\omega_{0}}{\omega_{\mathrm{c}}} 
\ll \left(
a_{\mathrm{e}}
\frac{\omega_{0}}{\omega_{\mathrm{c}}}
\right)_{\mathrm{trans}}^{\mathrm{coh}}, 
\\[2ex]
\sqrt{3}\left(\frac{a_{\mathrm{e}}^{2} \omega_{\mathrm{p}}^{2} \omega_{0}}{2}\right)^{\frac{1}{3}}
\left(\frac{\omega_{0}}{\omega_{\mathrm{c}}}\right)^{\frac{4}{3}}, 
\\
\quad \left(
a_{\mathrm{e}}
\frac{\omega_{0}}{\omega_{\mathrm{c}}}
\right)_{\mathrm{trans}}^{\mathrm{coh}}
\ll a_{\mathrm{e}} \frac{\omega_{0}}{\omega_{\mathrm{c}}} \ll 1,
\end{cases}
\label{eq:growth_rate_neutral_matome_Brillouin_letter}
\end{equation}
where the upper expression corresponds to ICS, and the lower one to SBS. The transition point is
\begin{equation}
    \left(
a_{\mathrm{e}}
\dfrac{\omega_{0}}{\omega_{\mathrm{c}}}
\right)_{\mathrm{trans}}^{\mathrm{coh}}\simeq4.4~\dfrac{\omega_{\text{c}}}{\omega_{\mathrm{p}}}
    \left(\dfrac{k_{\mathrm{B}}T_{\mathrm{e}}}{m_{\mathrm{e}} c^{2}}\right)^{\frac{3}{4}}
    \left(1+\dfrac{\omega_{\text{p}}^2}{\omega_{\text{c}}^2}\right)^{\frac{3}{4}}.
    \label{eq:transition_point_for_neutral_mode_letter}
\end{equation}

\subsection{Neutral mode: Linear Growth Rates for a Broadband Incident Wave}

For a broadband incident wave with bandwidth $\Delta\omega$, the maximum linear growth rate of the neutral mode is (see Eq. (130) in \citep{2025arXiv251012869N})
\begin{equation}
\begin{aligned}
\left(t_{\mathrm{neutral}}^{\mathrm{broad}}\right)^{-1}
\sim
\begin{cases}
&\pi
\left(
a_{\mathrm{e}}
\frac{\omega_{\mathrm{p}}}{\omega_{0}}
\right)^{2}
\left(
\frac{\omega_{0}}{\omega_{\mathrm{c}}}
\right)^{4}
\left(
\frac{\omega_{0}}{\Delta\omega}
\right)^{2}\!
\omega_{0},\\
&\quad
a_{\mathrm{e}}
\frac{\omega_{0}}{\omega_{\mathrm{c}}}
\ll
\left(
a_{\mathrm{e}}
\frac{\omega_{0}}{\omega_{\mathrm{c}}}
\right)_{\mathrm{trans}}^{\mathrm{broad}},\\[2.0ex]
&\frac{3}{2^{\frac{2}{3}}}
\left(
a_{\mathrm{e}}
\frac{\omega_{\mathrm{p}}}{\omega_{0}}
\right)^{\frac{4}{3}}
\left(
\frac{\omega_{0}}{\omega_{\mathrm{c}}}
\right)^{\frac{8}{3}}
\frac{\omega_{0}}{\Delta\omega}\,
\omega_{0},\\
&\quad
\left(
a_{\mathrm{e}}
\frac{\omega_{0}}{\omega_{\mathrm{c}}}
\right)_{\mathrm{trans}}^{\mathrm{broad}}
\ll
a_{\mathrm{e}}
\frac{\omega_{0}}{\omega_{\mathrm{c}}}
\ll1,
\end{cases}
\end{aligned}
\label{eq:neutral_summary_letter}
\end{equation}
where the upper expression corresponds to ICS, and the lower one to SBS. The transition point is
\begin{equation}
\left(a_{\mathrm{e}} \dfrac{\omega_0}{\omega_{\mathrm{c}}}\right)_{\mathrm{trans}}^{\mathrm{broad}} 
\sim 4.7 \times 10^{-1} 
\dfrac{\omega_{\mathrm{c}}}{\omega_{\mathrm{p}}} 
\left(\dfrac{\Delta \omega}{\omega_0}\right)^{\frac{3}{2}}.
\label{eq:weak_strong_transition_inc_neutral_letter}
\end{equation}

\subsection{Charged mode: Linear Growth Rates for a Broadband Incident Wave}

In the charged mode, ICS with Debye screening yields the maximum linear growth rate (see Eq. (133) in \citep{2025arXiv251012869N})
\begin{equation}
\begin{aligned}
\left(t_{\mathrm{charged}}^{\mathrm{broad}}\right)^{-1}
&\sim
\dfrac{\pi}{4}
\left( \dfrac{\omega_{0}}{\omega_{\mathrm{c}}} \right)^{2}
\left( 1 + \dfrac{\omega_{\text{p}}^2}{\omega_{\text{c}}^2} \right)^2
\dfrac{\omega_{\mathrm{p}}^{2} a_{\mathrm{e}}^{2}}{\omega_{0}}
\\
&\qquad \times
\left( \dfrac{8 k_{\mathrm{B}} T_{\mathrm{e}}}{m_{\mathrm{e}} c^{2}} \right)^{2}
\left( \dfrac{\omega_{0}}{\omega_{\mathrm{p}}} \right)^{4}
\left( \dfrac{\omega_{0}}{\Delta \omega} \right)^{2},
\end{aligned}
\label{eq:Compton_charged_Brillouin_growth_rate_Debye_letter}
\end{equation}
which can be applied in the regime $\omega_0 \ll \omega_{\mathrm p} \ll \omega_{\mathrm c}$.

\section{Spatially Resolved Nonlinear Evolution of Induced Scattering}
\label{app:spatially_resolved}
\begin{figure*}
\centering
\includegraphics[width=\textwidth]{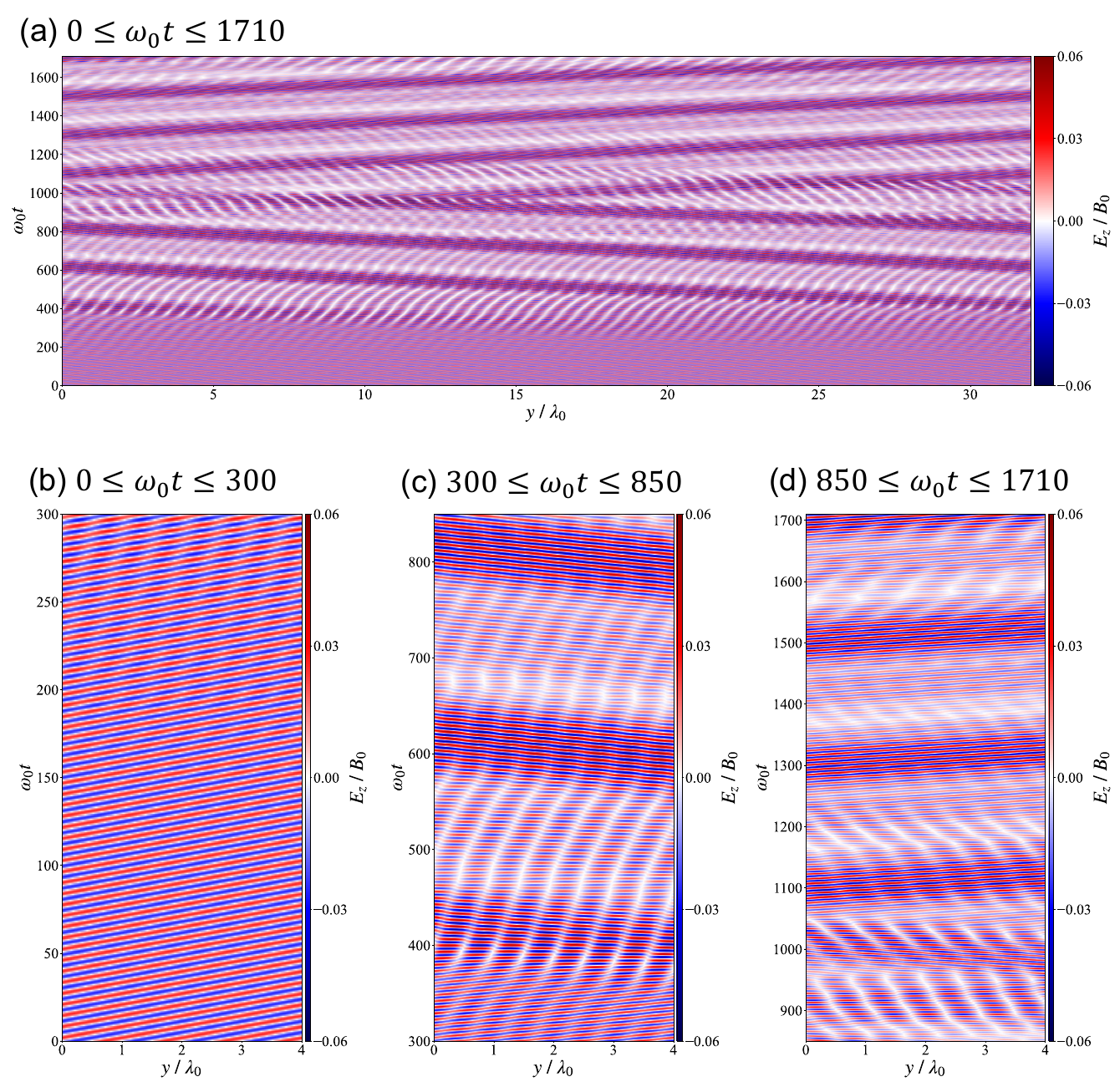}
\caption{\justifying \ioka{Time evolution of the spatial distribution of $E_z$ of the EM waves, for the run with the incident wave amplitude $\eta_{\mathrm{inci}}^{\mathrm{circ}}=0.0316$. The horizontal axis shows the spatial coordinate parallel to the background magnetic field, normalized by the incident wavelength, $y/\lambda_0$, and the vertical axis shows the normalized time, $\omega_0 t$. The color scale represents $E_z/B_0$, with red and blue corresponding to positive and negative values, respectively. Panel (a) shows the evolution over the full simulation time. Panels (b)--(d) show enlarged views of the region $0\leq y/\lambda_0\leq4$. They correspond to (b) the early stage dominated by the incident wave, (c) the stage in which energy is transferred to the backward-propagating scattered wave, and (d) the late stage in which the forward-propagating granddaughter wave grows again and the nonlinear energy exchange continues.}
}
\label{fig:spatial_entire}
\end{figure*}
\begin{figure}
\centering
\includegraphics[width=\columnwidth]{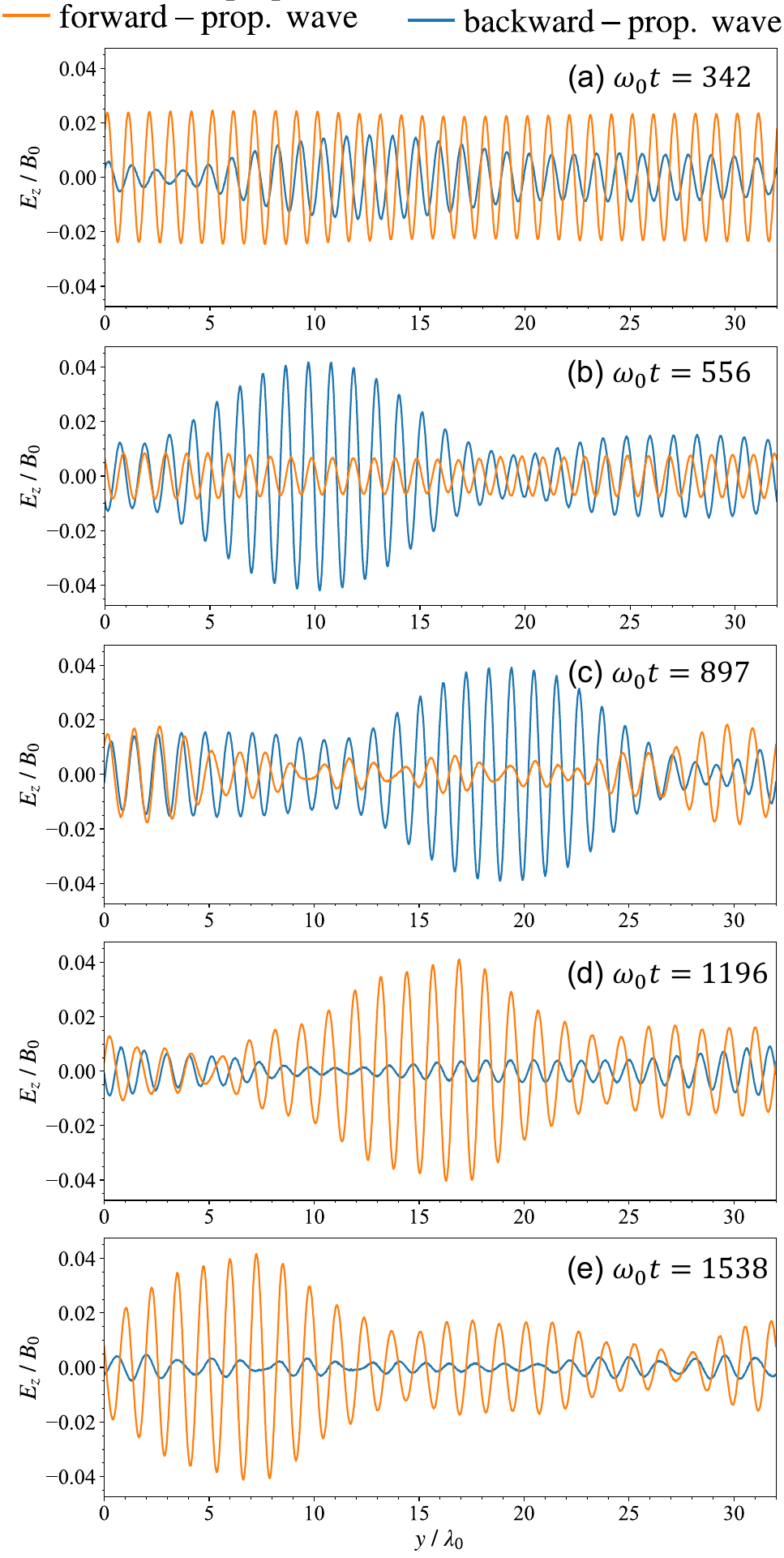}
\caption{\justifying \ioka{Spatial distributions of the electric field component $E_z$ of the forward-propagating component (orange) and the backward-propagating component (blue) of the EM wave at representative times. The horizontal axis shows the spatial coordinate parallel to the background magnetic field, normalized by the wavelength of the incident wave, $y/\lambda_0$, and the vertical axis shows $E_z/B_0$. Panel (a) shows the early stage in which the backward-propagating scattered wave grows linearly. Panel (b) shows the stage in which the scattered wave has grown substantially and the incident wave is attenuated. Panel (c) shows the stage in which the backward-propagating scattered wave decreases and the forward-propagating granddaughter wave begins to grow. Panel (d) shows the stage in which the granddaughter wave becomes dominant. Panel (e) shows the late nonlinear stage in which the backward-propagating component begins to grow again.}
}
\label{fig:spatial_figure}
\end{figure}
\ioka{To clarify the spatial structure associated with the nonlinear evolution of induced scattering shown in Fig.~\ref{fig:energy_distribution_letter}, we present the spatially resolved evolution of the EM wave. Fig.~\ref{fig:spatial_entire} shows the time evolution of the spatial distribution of the electric field component $E_z$. In panel (b), the incident wave initially propagates in the positive $y$ direction and appears as stripes with a positive slope. In panel (c), the backward-propagating scattered wave becomes dominant and appears as stripes with a negative slope. At later times, as shown in panel (d), the propagation direction and amplitude continue to change as the energy is repeatedly exchanged between the forward- and backward-propagating waves.}

\ioka{To examine this evolution in more detail, Fig.~\ref{fig:spatial_figure} shows the spatial distributions of the forward- and backward-propagating components of the EM wave at five representative times. We obtain these components by Fourier transforming the EM fields, separating the forward- and backward-propagating modes in Fourier space, and then applying the inverse Fourier transform to each component. The details of the Fourier decomposition are described in Sec.~IV B of \citet{2026arXiv260101169K}.}

\section{Spin period dependence of the regime maps}
\label{app:spin_period}
\begin{figure*}
\centering
\includegraphics[width=\textwidth]{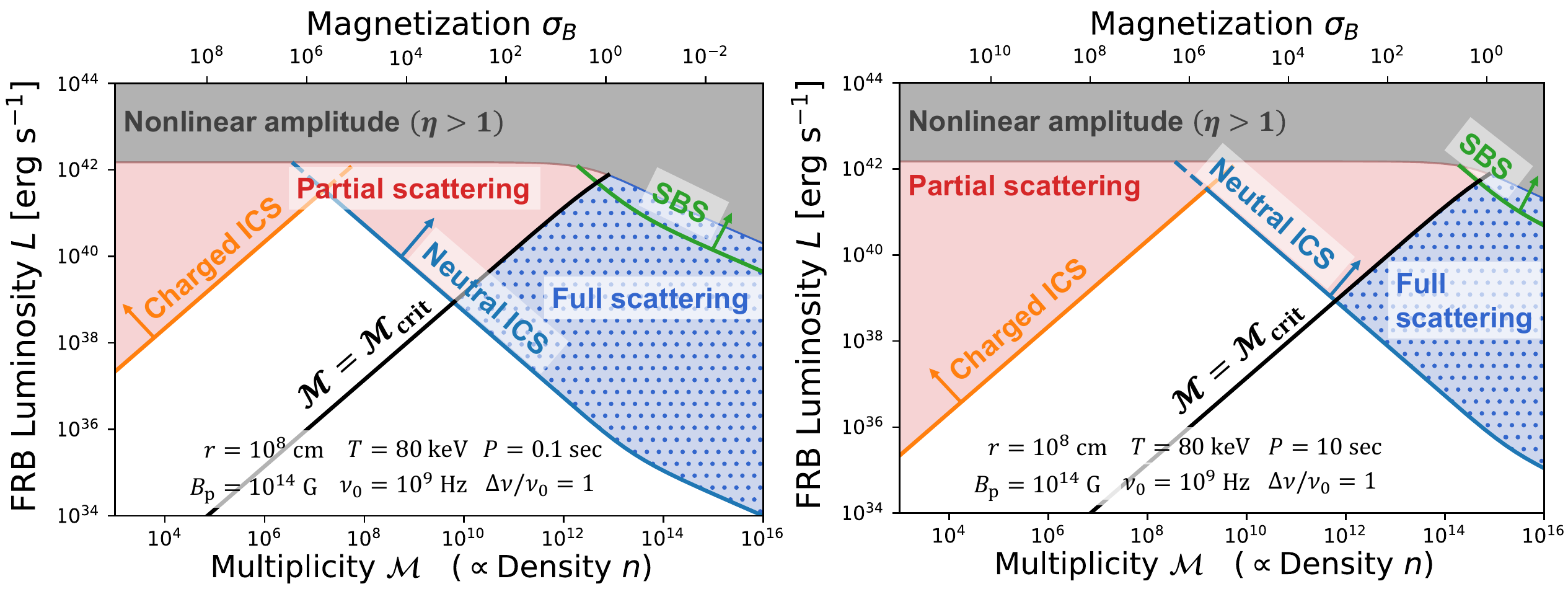}
\caption{\justifying Same as Fig.~\ref{Application_FRB_letter}, but for $P=0.1\,{\rm s}$ (left) and $P=10\,{\rm s}$ (right). The maps are evaluated at $r=10^8\,{\rm cm}$ with the other fiducial parameters fixed. The boundary between partial and full scattering shifts approximately as $\mathcal{M}_{\rm crit}\propto P$ from Eq. \eqref{eq:critical_multiplicity}.
}
\label{fig:LMmap_sec}
\end{figure*}
\begin{figure*}
\centering
\includegraphics[width=\textwidth]{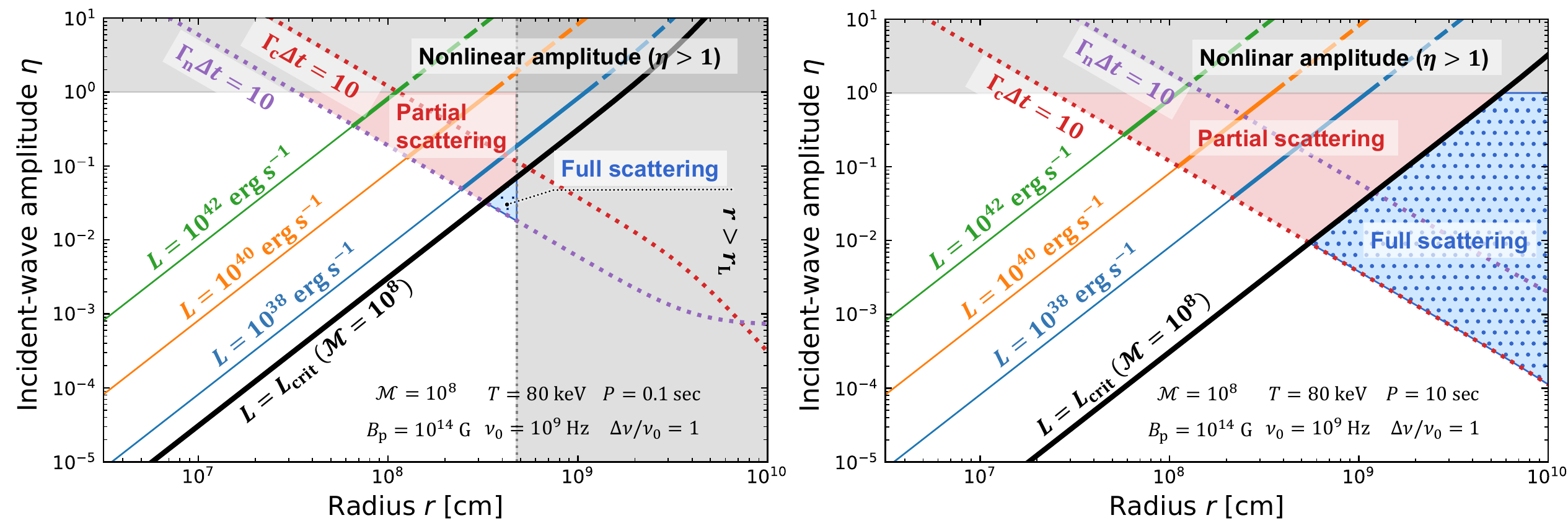}
\caption{\justifying Same as Fig.~\ref{radial_dependence}, but for $P=0.1\,{\rm s}$ and $P=10\,{\rm s}$. The light cylinder radius changes in proportion to $P$, while the linear growth thresholds move only weakly with $P$ from Eq. \eqref{eq:induced_scattering_FRB_letter}.
}
\label{fig:radial_sec}
\end{figure*}
In the main text, we adopt $P=1\,{\rm s}$ as the fiducial spin period. Here we show the same $L$--$\mathcal{M}$ regime maps in Fig. \ref{fig:LMmap_sec} and radial dependence maps in Fig. \ref{fig:radial_sec} for $P=0.1\,{\rm s}$ and $P=10\,{\rm s}$. The spin period changes the light cylinder radius as $r_{\rm L}=c/\Omega\simeq 4.8\times10^9\,{\rm cm}\,(P/1\,{\rm s})$ and shifts the plasma density through the Goldreich--Julian density.

% --- References ---
\nocite{*}
\bibliographystyle{apsrev4-2}
\bibliography{apssamp_cited_only}

@ARTICLE{2007Sci...318..777L,
       author = {{Lorimer}, D.~R. and {Bailes}, M. and {McLaughlin}, M.~A. and {Narkevic}, D.~J. and {Crawford}, F.},
        title = "{A Bright Millisecond Radio Burst of Extragalactic Origin}",
      journal = {Science},
         year = 2007,
        month = nov,
       volume = {318},
       number = {5851},
        pages = {777},
          doi = {10.1126/science.1147532},
archivePrefix = {arXiv},
       eprint = {0709.4301},
 primaryClass = {astro-ph},
       adsurl = {https://ui.adsabs.harvard.edu/abs/2007Sci...318..777L}
}

@ARTICLE{2020Natur.587...54C,
       author = {{CHIME/FRB Collaboration} and {Andersen}, B.~C. and {Bandura}, K.~M. and {Bhardwaj}, M. and {Bij}, A. and {Boyce}, M.~M. and {Boyle}, P.~J. and {Brar}, C. and {Cassanelli}, T. and {Chawla}, P. and {Chen}, T. and {Cliche}, J. -F. and {Cook}, A. and {Cubranic}, D. and {Curtin}, A.~P. and {Denman}, N.~T. and {Dobbs}, M. and {Dong}, F.~Q. and {Fandino}, M. and {Fonseca}, E. and {Gaensler}, B.~M. and {Giri}, U. and {Good}, D.~C. and {Halpern}, M. and {Hill}, A.~S. and {Hinshaw}, G.~F. and {H{\"o}fer}, C. and {Josephy}, A. and {Kania}, J.~W. and {Kaspi}, V.~M. and {Landecker}, T.~L. and {Leung}, C. and {Li}, D.~Z. and {Lin}, H. -H. and {Masui}, K.~W. and {McKinven}, R. and {Mena-Parra}, J. and {Merryfield}, M. and {Meyers}, B.~W. and {Michilli}, D. and {Milutinovic}, N. and {Mirhosseini}, A. and {M{\"u}nchmeyer}, M. and {Naidu}, A. and {Newburgh}, L.~B. and {Ng}, C. and {Patel}, C. and {Pen}, U. -L. and {Pinsonneault-Marotte}, T. and {Pleunis}, Z. and {Quine}, B.~M. and {Rafiei-Ravandi}, M. and {Rahman}, M. and {Ransom}, S.~M. and {Renard}, A. and {Sanghavi}, P. and {Scholz}, P. and {Shaw}, J.~R. and {Shin}, K. and {Siegel}, S.~R. and {Singh}, S. and {Smegal}, R.~J. and {Smith}, K.~M. and {Stairs}, I.~H. and {Tan}, C.~M. and {Tendulkar}, S.~P. and {Tretyakov}, I. and {Vanderlinde}, K. and {Wang}, H. and {Wulf}, D. and {Zwaniga}, A.~V.},
        title = "{A bright millisecond-duration radio burst from a Galactic magnetar}",
      journal = {\nat},
         year = 2020,
        month = nov,
       volume = {587},
       number = {7832},
        pages = {54-58},
          doi = {10.1038/s41586-020-2863-y},
archivePrefix = {arXiv},
       eprint = {2005.10324},
 primaryClass = {astro-ph.HE},
       adsurl = {https://ui.adsabs.harvard.edu/abs/2020Natur.587...54C}
}

@ARTICLE{2020Natur.587...59B,
       author = {{Bochenek}, C.~D. and {Ravi}, V. and {Belov}, K.~V. and {Hallinan}, G. and {Kocz}, J. and {Kulkarni}, S.~R. and {McKenna}, D.~L.},
        title = "{A fast radio burst associated with a Galactic magnetar}",
      journal = {\nat},
         year = 2020,
        month = nov,
       volume = {587},
       number = {7832},
        pages = {59-62},
          doi = {10.1038/s41586-020-2872-x},
archivePrefix = {arXiv},
       eprint = {2005.10828},
 primaryClass = {astro-ph.HE},
       adsurl = {https://ui.adsabs.harvard.edu/abs/2020Natur.587...59B}
}

@ARTICLE{2020ApJ...898L..29M,
       author = {{Mereghetti}, S. and {Savchenko}, V. and {Ferrigno}, C. and {G{\"o}tz}, D. and {Rigoselli}, M. and {Tiengo}, A. and {Bazzano}, A. and {Bozzo}, E. and {Coleiro}, A. and {Courvoisier}, T.~J. -L. and {Doyle}, M. and {Goldwurm}, A. and {Hanlon}, L. and {Jourdain}, E. and {von Kienlin}, A. and {Lutovinov}, A. and {Martin-Carrillo}, A. and {Molkov}, S. and {Natalucci}, L. and {Onori}, F. and {Panessa}, F. and {Rodi}, J. and {Rodriguez}, J. and {S{\'a}nchez-Fern{\'a}ndez}, C. and {Sunyaev}, R. and {Ubertini}, P.},
        title = "{INTEGRAL Discovery of a Burst with Associated Radio Emission from the Magnetar SGR 1935+2154}",
      journal = {\apjl},
         year = 2020,
        month = aug,
       volume = {898},
       number = {2},
          eid = {L29},
        pages = {L29},
          doi = {10.3847/2041-8213/aba2cf},
archivePrefix = {arXiv},
       eprint = {2005.06335},
 primaryClass = {astro-ph.HE},
       adsurl = {https://ui.adsabs.harvard.edu/abs/2020ApJ...898L..29M}
}

@ARTICLE{2008ApJ...682.1443L,
       author = {{Lyubarsky}, Yuri},
        title = "{Induced Scattering of Short Radio Pulses}",
      journal = {\apj},
         year = 2008,
        month = aug,
       volume = {682},
       number = {2},
        pages = {1443-1449},
          doi = {10.1086/589435},
archivePrefix = {arXiv},
       eprint = {0804.2069},
 primaryClass = {astro-ph},
       adsurl = {https://ui.adsabs.harvard.edu/abs/2008ApJ...682.1443L}
}

@ARTICLE{2023MNRAS.522.2133I,
       author = {{Iwamoto}, Masanori and {Sobacchi}, Emanuele and {Sironi}, Lorenzo},
        title = "{Kinetic simulations of the filamentation instability in pair plasmas}",
      journal = {\mnras},
         year = 2023,
        month = jun,
       volume = {522},
       number = {2},
        pages = {2133-2144},
          doi = {10.1093/mnras/stad1100},
archivePrefix = {arXiv},
       eprint = {2304.03577},
 primaryClass = {astro-ph.HE},
       adsurl = {https://ui.adsabs.harvard.edu/abs/2023MNRAS.522.2133I}
}

@ARTICLE{2024PhRvE.110a5205I,
       author = {{Ishizaki}, Wataru and {Ioka}, Kunihito},
        title = "{Parametric decay instability of circularly polarized Alfv{\'e}n waves in magnetically dominated plasma}",
      journal = {\pre},
         year = 2024,
        month = jul,
       volume = {110},
       number = {1},
          eid = {015205},
        pages = {015205},
          doi = {10.1103/PhysRevE.110.015205},
archivePrefix = {arXiv},
       eprint = {2404.15689},
 primaryClass = {astro-ph.HE},
       adsurl = {https://ui.adsabs.harvard.edu/abs/2024PhRvE.110a5205I}
}

@ARTICLE{1973PhFl...16.1480M,
       author = {{Max}, C.~E.},
        title = "{Parametric instability of a relativistically strong electromagnetic wave.}",
      journal = {Physics of Fluids},
         year = 1973,
        month = jan,
       volume = {16},
        pages = {1480-1489},
          doi = {10.1063/1.1694545},
       adsurl = {https://ui.adsabs.harvard.edu/abs/1973PhFl...16.1480M}
}

@ARTICLE{1976MNRAS.174...59B,
       author = {{Blandford}, R.~D. and {Scharlemann}, E.~T.},
        title = "{On the scattering and absorption of electromagnetic radiation with pulsar magnetospheres.}",
      journal = {\mnras},
         year = 1976,
        month = jan,
       volume = {174},
        pages = {59-85},
          doi = {10.1093/mnras/174.1.59},
       adsurl = {https://ui.adsabs.harvard.edu/abs/1976MNRAS.174...59B}
}

@ARTICLE{1978MNRAS.185..297W,
       author = {{Wilson}, D.~B. and {Rees}, M.~J.},
        title = "{Induced Compton scattering in pulsar winds}",
      journal = {\mnras},
         year = 1978,
        month = oct,
       volume = {185},
        pages = {297},
          doi = {10.1093/mnras/185.2.297},
       adsurl = {https://ui.adsabs.harvard.edu/abs/1978MNRAS.185..297W}
}

@ARTICLE{1982MNRAS.200..881W,
       author = {{Wilson}, D.~B.},
        title = "{Induced compton scattering in radiative transfer}",
      journal = {\mnras},
         year = 1982,
        month = sep,
       volume = {200},
        pages = {881-906},
          doi = {10.1093/mnras/200.4.881},
       adsurl = {https://ui.adsabs.harvard.edu/abs/1982MNRAS.200..881W}
}

@ARTICLE{1996AstL...22..399L,
       author = {{Lyubarskii}, Yu. E. and {Petrova}, S.~A.},
        title = "{Stimulated scattering of radio emission in pulsar magnetospheres}",
      journal = {Astronomy Letters},
         year = 1996,
        month = may,
       volume = {22},
       number = {3},
        pages = {399-408},
       adsurl = {https://ui.adsabs.harvard.edu/abs/1996AstL...22..399L}
}

@ARTICLE{1973PhFl...16.1522K,
       author = {{Kaw}, P. and {Schmidt}, G. and {Wilcox}, T.},
        title = "{Filamentation and trapping of electromagnetic radiation in plasmas}",
      journal = {Physics of Fluids},
         year = 1973,
        month = sep,
       volume = {16},
       number = {9},
        pages = {1522-1525},
          doi = {10.1063/1.1694552},
       adsurl = {https://ui.adsabs.harvard.edu/abs/1973PhFl...16.1522K}
}

@ARTICLE{1974PhRvL..33..209M,
       author = {{Max}, Claire Ellen and {Arons}, Jonathan and {Langdon}, A. Bruce},
        title = "{Self-Modulation and Self-Focusing of Electromagnetic Waves in Plasmas}",
      journal = {\prl},
         year = 1974,
        month = jul,
       volume = {33},
       number = {4},
        pages = {209-212},
          doi = {10.1103/PhysRevLett.33.209},
       adsurl = {https://ui.adsabs.harvard.edu/abs/1974PhRvL..33..209M}
}

@ARTICLE{1975PhFl...18.1002F,
       author = {{Forslund}, D.~W. and {Kindel}, J.~M. and {Lindman}, E.~L.},
        title = "{Theory of stimulated scattering processes in laser-irradiated plasmas}",
      journal = {Physics of Fluids},
         year = 1975,
        month = aug,
       volume = {18},
       number = {8},
        pages = {1002-1016},
          doi = {10.1063/1.861248},
       adsurl = {https://ui.adsabs.harvard.edu/abs/1975PhFl...18.1002F}
}

@ARTICLE{1994PhPl....1.1626T,
       author = {{Tabak}, Max and {Hammer}, James and {Glinsky}, Michael E. and {Kruer}, William L. and {Wilks}, Scott C. and {Woodworth}, John and {Campbell}, E. Michael and {Perry}, Michael D. and {Mason}, Rodney J.},
        title = "{Ignition and high gain with ultrapowerful lasers*}",
      journal = {Physics of Plasmas},
         year = 1994,
        month = may,
       volume = {1},
       number = {5},
        pages = {1626-1634},
          doi = {10.1063/1.870664},
       adsurl = {https://ui.adsabs.harvard.edu/abs/1994PhPl....1.1626T}
}

@ARTICLE{1996PhRvL..77.2483D,
       author = {{Deutsch}, C. and {Furukawa}, H. and {Mima}, K. and {Murakami}, M. and {Nishihara}, K.},
        title = "{Interaction Physics of the Fast Ignitor Concept}",
      journal = {\prl},
         year = 1996,
        month = sep,
       volume = {77},
       number = {12},
        pages = {2483-2486},
          doi = {10.1103/PhysRevLett.77.2483},
       adsurl = {https://ui.adsabs.harvard.edu/abs/1996PhRvL..77.2483D}
}

@ARTICLE{1979PhFl...22.1089K,
       author = {{Kwan}, T. and {Dawson}, J.~M.},
        title = "{Investigation of the free electron laser with a guide magnetic field}",
      journal = {Physics of Fluids},
         year = 1979,
        month = jun,
       volume = {22},
       number = {6},
        pages = {1089-1103},
          doi = {10.1063/1.862702},
       adsurl = {https://ui.adsabs.harvard.edu/abs/1979PhFl...22.1089K}
}

@ARTICLE{1980PhFl...23.2376F,
       author = {{Friedland}, L.},
        title = "{Electron beam dynamics in combined guide and pump magnetic fields for free electron laser applications}",
      journal = {Physics of Fluids},
         year = 1980,
        month = dec,
       volume = {23},
       number = {12},
        pages = {2376-2382},
          doi = {10.1063/1.862942},
       adsurl = {https://ui.adsabs.harvard.edu/abs/1980PhFl...23.2376F}
}

@ARTICLE{2021ApJ...922L...7B,
       author = {{Beloborodov}, Andrei M.},
        title = "{Can a Strong Radio Burst Escape the Magnetosphere of a Magnetar?}",
      journal = {\apjl},
         year = 2021,
        month = nov,
       volume = {922},
       number = {1},
          eid = {L7},
        pages = {L7},
          doi = {10.3847/2041-8213/ac2fa0},
archivePrefix = {arXiv},
       eprint = {2108.07881},
 primaryClass = {astro-ph.HE},
       adsurl = {https://ui.adsabs.harvard.edu/abs/2021ApJ...922L...7B}
}

@ARTICLE{2022PhRvL.128y5003B,
       author = {{Beloborodov}, Andrei M.},
        title = "{Scattering of Ultrastrong Electromagnetic Waves by Magnetized Particles}",
      journal = {\prl},
         year = 2022,
        month = jun,
       volume = {128},
       number = {25},
          eid = {255003},
        pages = {255003},
          doi = {10.1103/PhysRevLett.128.255003},
archivePrefix = {arXiv},
       eprint = {2108.05464},
 primaryClass = {astro-ph.HE},
       adsurl = {https://ui.adsabs.harvard.edu/abs/2022PhRvL.128y5003B}
}

@ARTICLE{2022MNRAS.515.2020Q,
       author = {{Qu}, Yuanhong and {Kumar}, Pawan and {Zhang}, Bing},
        title = "{Transparency of fast radio burst waves in magnetar magnetospheres}",
      journal = {\mnras},
         year = 2022,
        month = sep,
       volume = {515},
       number = {2},
        pages = {2020-2031},
          doi = {10.1093/mnras/stac1910},
archivePrefix = {arXiv},
       eprint = {2204.10953},
 primaryClass = {astro-ph.HE},
       adsurl = {https://ui.adsabs.harvard.edu/abs/2022MNRAS.515.2020Q}
}

@ARTICLE{2022arXiv221013506C,
       author = {{Chen}, Alexander Y. and {Yuan}, Yajie and {Li}, Xinyu and {Mahlmann}, Jens F.},
        title = "{Propagation of a Strong Fast Magnetosonic Wave in the Magnetosphere of a Neutron Star}",
      journal = {arXiv e-prints},
         year = 2022,
        month = oct,
          eid = {arXiv:2210.13506},
        pages = {arXiv:2210.13506},
          doi = {10.48550/arXiv.2210.13506},
archivePrefix = {arXiv},
       eprint = {2210.13506},
 primaryClass = {astro-ph.HE},
       adsurl = {https://ui.adsabs.harvard.edu/abs/2022arXiv221013506C}
}

@ARTICLE{2023ApJ...959...34B,
       author = {{Beloborodov}, Andrei M.},
        title = "{Monster Radiative Shocks in the Perturbed Magnetospheres of Neutron Stars}",
      journal = {\apj},
         year = 2023,
        month = dec,
       volume = {959},
       number = {1},
          eid = {34},
        pages = {34},
          doi = {10.3847/1538-4357/acf659},
archivePrefix = {arXiv},
       eprint = {2210.13509},
 primaryClass = {astro-ph.HE},
       adsurl = {https://ui.adsabs.harvard.edu/abs/2023ApJ...959...34B}
}

@ARTICLE{2024ApJ...975..223B,
       author = {{Beloborodov}, Andrei M.},
        title = "{Damping of Strong GHz Waves near Magnetars and the Origin of Fast Radio Bursts}",
      journal = {\apj},
         year = 2024,
        month = nov,
       volume = {975},
       number = {2},
          eid = {223},
        pages = {223},
          doi = {10.3847/1538-4357/ad698c},
archivePrefix = {arXiv},
       eprint = {2307.12182},
 primaryClass = {astro-ph.HE},
       adsurl = {https://ui.adsabs.harvard.edu/abs/2024ApJ...975..223B}
}

@ARTICLE{2024ApJ...975..226H,
       author = {{Huang}, Yu-Chen and {Dai}, Zi-Gao},
        title = "{Fast Radio Bursts with Narrow Beaming Angles Can Escape from Magnetar Magnetospheres}",
      journal = {\apj},
         year = 2024,
        month = nov,
       volume = {975},
       number = {2},
          eid = {226},
        pages = {226},
          doi = {10.3847/1538-4357/ad822e},
archivePrefix = {arXiv},
       eprint = {2410.04065},
 primaryClass = {astro-ph.HE},
       adsurl = {https://ui.adsabs.harvard.edu/abs/2024ApJ...975..226H}
}

@ARTICLE{2024PhRvD.109d3048N,
       author = {{Nishiura}, Rei and {Ioka}, Kunihito},
        title = "{Collective Thomson scattering in magnetized electron and positron pair plasma and the application to induced Compton scattering}",
      journal = {\prd},
         year = 2024,
        month = feb,
       volume = {109},
       number = {4},
          eid = {043048},
        pages = {043048},
          doi = {10.1103/PhysRevD.109.043048},
archivePrefix = {arXiv},
       eprint = {2310.02306},
 primaryClass = {astro-ph.HE},
       adsurl = {https://ui.adsabs.harvard.edu/abs/2024PhRvD.109d3048N}
}

@ARTICLE{2024A&A...690A.332S,
       author = {{Sobacchi}, E. and {Iwamoto}, M. and {Sironi}, L. and {Piran}, T.},
        title = "{Escape of fast radio bursts from magnetars}",
      journal = {\aap},
         year = 2024,
        month = oct,
       volume = {690},
          eid = {A332},
        pages = {A332},
          doi = {10.1051/0004-6361/202451725},
archivePrefix = {arXiv},
       eprint = {2409.10732},
 primaryClass = {astro-ph.HE},
       adsurl = {https://ui.adsabs.harvard.edu/abs/2024A&A...690A.332S}
}

@ARTICLE{2025PhRvD.111f3055N,
       author = {{Nishiura}, Rei and {Kamijima}, Shoma F. and {Iwamoto}, Masanori and {Ioka}, Kunihito},
        title = "{Induced Compton scattering in magnetized electron and positron pair plasma}",
      journal = {\prd},
         year = 2025,
        month = mar,
       volume = {111},
       number = {6},
          eid = {063055},
        pages = {063055},
          doi = {10.1103/PhysRevD.111.063055},
archivePrefix = {arXiv},
       eprint = {2411.00936},
 primaryClass = {astro-ph.HE},
       adsurl = {https://ui.adsabs.harvard.edu/abs/2025PhRvD.111f3055N}
}

@ARTICLE{2022Natur.607..256C,
       author = {{CHIME/FRB Collaboration}, Andersen, Bridget C. and {Bandura}, Kevin and {Bhardwaj}, Mohit and {Boyle}, P.~J. and {Brar}, Charanjot and {Breitman}, Daniela and {Cassanelli}, Tomas and {Chatterjee}, Shami and {Chawla}, Pragya and {Cliche}, Jean-Fran{\c{c}}ois and {Cubranic}, Davor and {Curtin}, Alice P. and {Deng}, Meiling and {Dobbs}, Matt and {Dong}, Fengqiu Adam and {Fonseca}, Emmanuel and {Gaensler}, B.~M. and {Giri}, Utkarsh and {Good}, Deborah C. and {Hill}, Alex S. and {Josephy}, Alexander and {Kaczmarek}, J.~F. and {Kader}, Zarif and {Kania}, Joseph and {Kaspi}, Victoria M. and {Leung}, Calvin and {Li}, D.~Z. and {Lin}, Hsiu-Hsien and {Masui}, Kiyoshi W. and {McKinven}, Ryan and {Mena-Parra}, Juan and {Merryfield}, Marcus and {Meyers}, B.~W. and {Michilli}, D. and {Naidu}, Arun and {Newburgh}, Laura and {Ng}, C. and {Ordog}, Anna and {Patel}, Chitrang and {Pearlman}, Aaron B. and {Pen}, Ue-Li and {Petroff}, Emily and {Pleunis}, Ziggy and {Rafiei-Ravandi}, Masoud and {Rahman}, Mubdi and {Ransom}, Scott and {Renard}, Andre and {Sanghavi}, Pranav and {Scholz}, Paul and {Shaw}, J. Richard and {Shin}, Kaitlyn and {Siegel}, Seth R. and {Singh}, Saurabh and {Smith}, Kendrick and {Stairs}, Ingrid and {Tan}, Chia Min and {Tendulkar}, Shriharsh P. and {Vanderlinde}, Keith and {Wiebe}, D.~V. and {Wulf}, Dallas and {Zwaniga}, Andrew},
        title = "{Sub-second periodicity in a fast radio burst}",
      journal = {\nat},
         year = 2022,
        month = jul,
       volume = {607},
       number = {7918},
        pages = {256-259},
          doi = {10.1038/s41586-022-04841-8},
archivePrefix = {arXiv},
       eprint = {2107.08463},
 primaryClass = {astro-ph.HE},
       adsurl = {https://ui.adsabs.harvard.edu/abs/2022Natur.607..256C}
}

@ARTICLE{2022NatAs...6..393N,
       author = {{Nimmo}, K. and {Hessels}, J.~W.~T. and {Kirsten}, F. and {Keimpema}, A. and {Cordes}, J.~M. and {Snelders}, M.~P. and {Hewitt}, D.~M. and {Karuppusamy}, R. and {Archibald}, A.~M. and {Bezrukovs}, V. and {Bhardwaj}, M. and {Blaauw}, R. and {Buttaccio}, S.~T. and {Cassanelli}, T. and {Conway}, J.~E. and {Corongiu}, A. and {Feiler}, R. and {Fonseca}, E. and {Forss{\'e}n}, O. and {Gawro{\'n}ski}, M. and {Giroletti}, M. and {Kharinov}, M.~A. and {Leung}, C. and {Lindqvist}, M. and {Maccaferri}, G. and {Marcote}, B. and {Masui}, K.~W. and {Mckinven}, R. and {Melnikov}, A. and {Michilli}, D. and {Mikhailov}, A.~G. and {Ng}, C. and {Orbidans}, A. and {Ould-Boukattine}, O.~S. and {Paragi}, Z. and {Pearlman}, A.~B. and {Petroff}, E. and {Rahman}, M. and {Scholz}, P. and {Shin}, K. and {Smith}, K.~M. and {Stairs}, I.~H. and {Surcis}, G. and {Tendulkar}, S.~P. and {Vlemmings}, W. and {Wang}, N. and {Yang}, J. and {Yuan}, J.~P.},
        title = "{Burst timescales and luminosities as links between young pulsars and fast radio bursts}",
      journal = {Nature Astronomy},
         year = 2022,
        month = feb,
       volume = {6},
        pages = {393-401},
          doi = {10.1038/s41550-021-01569-9},
archivePrefix = {arXiv},
       eprint = {2105.11446},
 primaryClass = {astro-ph.HE},
       adsurl = {https://ui.adsabs.harvard.edu/abs/2022NatAs...6..393N}
}

@ARTICLE{2025Natur.637...48N,
       author = {{Nimmo}, Kenzie and {Pleunis}, Ziggy and {Beniamini}, Paz and {Kumar}, Pawan and {Lanman}, Adam E. and {Li}, D.~Z. and {Main}, Robert and {Sammons}, Mawson W. and {Andrew}, Shion and {Bhardwaj}, Mohit and {Chatterjee}, Shami and {Curtin}, Alice P. and {Fonseca}, Emmanuel and {Gaensler}, B.~M. and {Joseph}, Ronniy C. and {Kader}, Zarif and {Kaspi}, Victoria M. and {Lazda}, Mattias and {Leung}, Calvin and {Masui}, Kiyoshi W. and {Mckinven}, Ryan and {Michilli}, Daniele and {Pandhi}, Ayush and {Pearlman}, Aaron B. and {Rafiei-Ravandi}, Masoud and {Sand}, Ketan R. and {Shin}, Kaitlyn and {Smith}, Kendrick and {Stairs}, Ingrid H.},
        title = "{Magnetospheric origin of a fast radio burst constrained using scintillation}",
      journal = {\nat},
         year = 2025,
        month = jan,
       volume = {637},
       number = {8044},
        pages = {48-51},
          doi = {10.1038/s41586-024-08297-w},
archivePrefix = {arXiv},
       eprint = {2406.11053},
 primaryClass = {astro-ph.HE},
       adsurl = {https://ui.adsabs.harvard.edu/abs/2025Natur.637...48N}
}

@ARTICLE{1978A&A....66..139S,
       author = {{Sweeney}, G.~S.~S. and {Stewart}, P.},
        title = "{Magneto-parametric instabilities in the Crab Nebula: II.}",
      journal = {\aap},
         year = 1978,
        month = may,
       volume = {66},
       number = {1-2},
        pages = {139-153},
       adsurl = {https://ui.adsabs.harvard.edu/abs/1978A&A....66..139S}
}

@ARTICLE{1998PhRvE..57..994M,
       author = {{Mu{\~n}oz}, V. and {Gomberoff}, L.},
        title = "{Parametric decays of a circularly polarized electromagnetic wave in an electron-positron magnetized plasma}",
      journal = {\pre},
         year = 1998,
        month = jan,
       volume = {57},
       number = {1},
        pages = {994-1004},
          doi = {10.1103/PhysRevE.57.994},
       adsurl = {https://ui.adsabs.harvard.edu/abs/1998PhRvE..57..994M}
}

@ARTICLE{1999PhRvE..59.4552M,
       author = {{Machabeli}, G.~Z. and {Vladimirov}, S.~V. and {Melrose}, D.~B.},
        title = "{Nonlinear dynamics of an ordinary electromagnetic mode in a pair plasma}",
      journal = {\pre},
         year = 1999,
        month = apr,
       volume = {59},
       number = {4},
        pages = {4552-4558},
          doi = {10.1103/PhysRevE.59.4552},
       adsurl = {https://ui.adsabs.harvard.edu/abs/1999PhRvE..59.4552M}
}

@ARTICLE{2003PhRvE..67d6406M,
       author = {{Matsukiyo}, S. and {Hada}, T.},
        title = "{Parametric instabilities of circularly polarized Alfv{\'e}n waves in a relativistic electron-positron plasma}",
      journal = {\pre},
         year = 2003,
        month = apr,
       volume = {67},
       number = {4},
          eid = {046406},
        pages = {046406},
          doi = {10.1103/PhysRevE.67.046406},
       adsurl = {https://ui.adsabs.harvard.edu/abs/2003PhRvE..67d6406M}
}

@ARTICLE{2006EP&S...58.1213M,
       author = {{Mu{\~n}oz}, V. and {Hada}, T. and {Matsukiyo}, S.},
        title = "{Kinetic effects on the parametric decays of Alfv{\'e}n waves in relativistic pair plasmas}",
      journal = {Earth, Planets and Space},
         year = 2006,
        month = sep,
       volume = {58},
       number = {9},
        pages = {1213-1217},
          doi = {10.1186/BF03352012},
       adsurl = {https://ui.adsabs.harvard.edu/abs/2006EP&S...58.1213M}
}

@ARTICLE{2012PhPl...19h2104L,
       author = {{L{\'o}pez}, Rodrigo A. and {Asenjo}, Felipe A. and {Mu{\~n}oz}, V{\'\i}ctor and {Alejandro Valdivia}, J.},
        title = "{Parametric decays in relativistic magnetized electron-positron plasmas with relativistic temperatures}",
      journal = {Physics of Plasmas},
         year = 2012,
        month = aug,
       volume = {19},
       number = {8},
          eid = {082104},
        pages = {082104},
          doi = {10.1063/1.4742315},
       adsurl = {https://ui.adsabs.harvard.edu/abs/2012PhPl...19h2104L}
}

@ARTICLE{2014PhPl...21c2102L,
       author = {{L{\'o}pez}, Rodrigo A. and {Mu{\~n}oz}, V{\'\i}ctor and {Vi{\~n}as}, Adolfo F. and {Alejandro Valdivia}, J.},
        title = "{Particle-in-cell simulation for parametric decays of a circularly polarized Alfv{\'e}n wave in relativistic thermal electron-positron plasma}",
      journal = {Physics of Plasmas},
         year = 2014,
        month = mar,
       volume = {21},
       number = {3},
          eid = {032102},
        pages = {032102},
          doi = {10.1063/1.4867255},
       adsurl = {https://ui.adsabs.harvard.edu/abs/2014PhPl...21c2102L}
}

@ARTICLE{2014NPGeo..21..217M,
       author = {{Mu{\~n}oz}, V. and {Asenjo}, F.~A. and {Dom{\'\i}nguez}, M. and {L{\'o}pez}, R.~A. and {Valdivia}, J.~A. and {Vi{\~n}as}, A. and {Hada}, T.},
        title = "{Large-amplitude electromagnetic waves in magnetized relativistic plasmas with temperature}",
      journal = {Nonlinear Processes in Geophysics},
         year = 2014,
        month = feb,
       volume = {21},
       number = {1},
        pages = {217-236},
          doi = {10.5194/npg-21-217-2014},
       adsurl = {https://ui.adsabs.harvard.edu/abs/2014NPGeo..21..217M}
}

@ARTICLE{2025arXiv251012869N,
       author = {{Nishiura}, Rei and {Kamijima}, Shoma F. and {Ioka}, Kunihito},
        title = "{Unified kinetic theory of induced scattering: Compton, Brillouin, and Raman processes in magnetized electron and positron pair plasma}",
      journal = {arXiv e-prints},
         year = 2025,
        month = oct,
          eid = {arXiv:2510.12869},
        pages = {arXiv:2510.12869},
          doi = {10.48550/arXiv.2510.12869},
archivePrefix = {arXiv},
       eprint = {2510.12869},
 primaryClass = {astro-ph.HE},
       adsurl = {https://ui.adsabs.harvard.edu/abs/2025arXiv251012869N}
}

@BOOK{1969npt..book.....S,
       author = {{Sagdeev}, R.~Z. and {Galeev}, A.~A.},
        title = "{Nonlinear Plasma Theory}",
         year = 1969,
       adsurl = {https://ui.adsabs.harvard.edu/abs/1969npt..book.....S}
}

@ARTICLE{2017RvMPP...1....5M,
       author = {{Melrose}, D.~B.},
        title = "{Coherent emission mechanisms in astrophysical plasmas}",
      journal = {Reviews of Modern Plasma Physics},
         year = 2017,
        month = dec,
       volume = {1},
       number = {1},
          eid = {5},
        pages = {5},
          doi = {10.1007/s41614-017-0007-0},
archivePrefix = {arXiv},
       eprint = {1707.02009},
 primaryClass = {physics.plasm-ph},
       adsurl = {https://ui.adsabs.harvard.edu/abs/2017RvMPP...1....5M}
}

@ARTICLE{2021ApJ...923....1P,
       author = {{Pleunis}, Ziggy and {Good}, Deborah C. and {Kaspi}, Victoria M. and {Mckinven}, Ryan and {Ransom}, Scott M. and {Scholz}, Paul and {Bandura}, Kevin and {Bhardwaj}, Mohit and {Boyle}, P.~J. and {Brar}, Charanjot and {Cassanelli}, Tomas and {Chawla}, Pragya and {(Adam) Dong}, Fengqiu and {Fonseca}, Emmanuel and {Gaensler}, B.~M. and {Josephy}, Alexander and {Kaczmarek}, Jane F. and {Leung}, Calvin and {Lin}, Hsiu-Hsien and {Masui}, Kiyoshi W. and {Mena-Parra}, Juan and {Michilli}, Daniele and {Ng}, Cherry and {Patel}, Chitrang and {Rafiei-Ravandi}, Masoud and {Rahman}, Mubdi and {Sanghavi}, Pranav and {Shin}, Kaitlyn and {Smith}, Kendrick M. and {Stairs}, Ingrid H. and {Tendulkar}, Shriharsh P.},
        title = "{Fast Radio Burst Morphology in the First CHIME/FRB Catalog}",
      journal = {\apj},
         year = 2021,
        month = dec,
       volume = {923},
       number = {1},
          eid = {1},
        pages = {1},
          doi = {10.3847/1538-4357/ac33ac},
archivePrefix = {arXiv},
       eprint = {2106.04356},
 primaryClass = {astro-ph.HE},
       adsurl = {https://ui.adsabs.harvard.edu/abs/2021ApJ...923....1P}
}

@ARTICLE{1974PhFl...17..849T,
       author = {{Thomson}, J.~J. and {Kruer}, W.~L. and {Bodner}, S.~E. and {DeGroot}, J.~S.},
        title = "{Parametric instability thresholds and their control}",
      journal = {Physics of Fluids},
         year = 1974,
        month = apr,
       volume = {17},
       number = {4},
        pages = {849-851},
          doi = {10.1063/1.1694799},
       adsurl = {https://ui.adsabs.harvard.edu/abs/1974PhFl...17..849T}
}

@ARTICLE{1994ApJ...422..304T,
       author = {{Thompson}, C. and {Blandford}, R.~D. and {Evans}, Charles R. and {Phinney}, E.~S.},
        title = "{Physical Processes in Eclipsing Pulsars: Eclipse Mechanisms and Diagnostics}",
      journal = {\apj},
         year = 1994,
        month = feb,
       volume = {422},
        pages = {304},
          doi = {10.1086/173728},
       adsurl = {https://ui.adsabs.harvard.edu/abs/1994ApJ...422..304T}
}

@book{kruer2019physics,
  title={The Physics Of Laser Plasma Interactions},
  author={Kruer, W.},
  isbn={9781000754469},
  url={https://books.google.co.jp/books?id=ideuDwAAQBAJ},
  year={2019},
  publisher={CRC Press}
}

@ARTICLE{2021PPCF...63i4003B,
       author = {{Brand{\~a}o}, B. and {Santos}, J.~E. and {Trines}, R.~M.~G.~M. and {Bingham}, R. and {Silva}, L.~O.},
        title = "{Bandwidth effects in stimulated Brillouin scattering driven by partially incoherent light}",
      journal = {Plasma Physics and Controlled Fusion},
         year = 2021,
        month = sep,
       volume = {63},
       number = {9},
          eid = {094003},
        pages = {094003},
          doi = {10.1088/1361-6587/ac11b5},
archivePrefix = {arXiv},
       eprint = {2104.04097},
 primaryClass = {physics.plasm-ph},
       adsurl = {https://ui.adsabs.harvard.edu/abs/2021PPCF...63i4003B}
}

@ARTICLE{2023RvMPP...7....1Z,
       author = {{Zhao}, Yao and {Weng}, Su-Ming and {Ma}, Hang-Hang and {Bai}, Xiao-Jun and {Sheng}, Zheng-Ming},
        title = "{Mitigation of laser plasma parametric instabilities with broadband lasers}",
      journal = {Reviews of Modern Plasma Physics},
         year = 2023,
        month = dec,
       volume = {7},
       number = {1},
          eid = {1},
        pages = {1},
          doi = {10.1007/s41614-022-00105-0},
       adsurl = {https://ui.adsabs.harvard.edu/abs/2023RvMPP...7....1Z}
}

@ARTICLE{2005Natur.434.1104G,
       author = {{Gaensler}, B.~M. and {Kouveliotou}, C. and {Gelfand}, J.~D. and {Taylor}, G.~B. and {Eichler}, D. and {Wijers}, R.~A.~M.~J. and {Granot}, J. and {Ramirez-Ruiz}, E. and {Lyubarsky}, Y.~E. and {Hunstead}, R.~W. and {Campbell-Wilson}, D. and {van der Horst}, A.~J. and {McLaughlin}, M.~A. and {Fender}, R.~P. and {Garrett}, M.~A. and {Newton-McGee}, K.~J. and {Palmer}, D.~M. and {Gehrels}, N. and {Woods}, P.~M.},
        title = "{An expanding radio nebula produced by a giant flare from the magnetar SGR 1806-20}",
      journal = {\nat},
         year = 2005,
        month = apr,
       volume = {434},
       number = {7037},
        pages = {1104-1106},
          doi = {10.1038/nature03498},
archivePrefix = {arXiv},
       eprint = {astro-ph/0502393},
 primaryClass = {astro-ph},
       adsurl = {https://ui.adsabs.harvard.edu/abs/2005Natur.434.1104G}
}

@ARTICLE{2005ApJ...634L..89G,
       author = {{Gelfand}, J.~D. and {Lyubarsky}, Y.~E. and {Eichler}, D. and {Gaensler}, B.~M. and {Taylor}, G.~B. and {Granot}, J. and {Newton-McGee}, K.~J. and {Ramirez-Ruiz}, E. and {Kouveliotou}, C. and {Wijers}, R.~A.~M.~J.},
        title = "{A Rebrightening of the Radio Nebula Associated with the 2004 December 27 Giant Flare from SGR 1806-20}",
      journal = {\apjl},
         year = 2005,
        month = nov,
       volume = {634},
       number = {1},
        pages = {L89-L92},
          doi = {10.1086/498643},
archivePrefix = {arXiv},
       eprint = {astro-ph/0503269},
 primaryClass = {astro-ph},
       adsurl = {https://ui.adsabs.harvard.edu/abs/2005ApJ...634L..89G}
}

@ARTICLE{2006ApJ...638..391G,
       author = {{Granot}, J. and {Ramirez-Ruiz}, E. and {Taylor}, G.~B. and {Eichler}, D. and {Lyubarsky}, Y.~E. and {Wijers}, R.~A.~M.~J. and {Gaensler}, B.~M. and {Gelfand}, J.~D. and {Kouveliotou}, C.},
        title = "{Diagnosing the Outflow from the SGR 1806-20 Giant Flare with Radio Observations}",
      journal = {\apj},
         year = 2006,
        month = feb,
       volume = {638},
       number = {1},
        pages = {391-396},
          doi = {10.1086/497680},
archivePrefix = {arXiv},
       eprint = {astro-ph/0503251},
 primaryClass = {astro-ph},
       adsurl = {https://ui.adsabs.harvard.edu/abs/2006ApJ...638..391G}
}

@ARTICLE{1968CzJPh..18.1280K,
       author = {{Kl{\'\i}ma}, R.},
        title = "{The drifts and hydrodynamics of particles in a field with a high-frequency component}",
      journal = {Czechoslovak Journal of Physics},
         year = 1968,
        month = oct,
       volume = {18},
       number = {10},
        pages = {1280-1291},
          doi = {10.1007/BF01690802},
       adsurl = {https://ui.adsabs.harvard.edu/abs/1968CzJPh..18.1280K}
}

@ARTICLE{1977PhRvL..39..402C,
       author = {{Cary}, J.~R. and {Kaufman}, A.~N.},
        title = "{Ponderomotive Force and Linear Susceptibility in Vlasov Plasma}",
      journal = {\prl},
         year = 1977,
        month = aug,
       volume = {39},
       number = {7},
        pages = {402-404},
          doi = {10.1103/PhysRevLett.39.402},
       adsurl = {https://ui.adsabs.harvard.edu/abs/1977PhRvL..39..402C}
}

@ARTICLE{1981PhFl...24.1238C,
       author = {{Cary}, J.~R. and {Kaufman}, A.~N.},
        title = "{Ponderomotive effects in collisionless plasma: A Lie transform approach}",
      journal = {Physics of Fluids},
         year = 1981,
        month = jul,
       volume = {24},
       number = {7},
        pages = {1238-1250},
          doi = {10.1063/1.863527},
       adsurl = {https://ui.adsabs.harvard.edu/abs/1981PhFl...24.1238C}
}

@ARTICLE{1981PhRvL..46..240H,
       author = {{Hatori}, T. and {Washimi}, H.},
        title = "{Covariant Form of the Ponderomotive Potentials in a Magnetized Plasma}",
      journal = {\prl},
         year = 1981,
        month = jan,
       volume = {46},
       number = {4},
        pages = {240-243},
          doi = {10.1103/PhysRevLett.46.240},
       adsurl = {https://ui.adsabs.harvard.edu/abs/1981PhRvL..46..240H}
}

@article{10.1063/1.864196,
    author = {Lee, Nam C. and Parks, G. K.},
    title = "{Ponderomotive force in a warm two‐fluid plasma}",
    journal = {The Physics of Fluids},
    volume = {26},
    number = {3},
    pages = {724-729},
    year = {1983},
    month = {03},
    issn = {0031-9171},
    doi = {10.1063/1.864196},
    url = {https://doi.org/10.1063/1.864196},
    eprint = {https://pubs.aip.org/aip/pfl/article-pdf/26/3/724/12747913/724\_1\_online.pdf},
}

@ARTICLE{1996GeoRL..23..327L,
       author = {{Lee}, Nam C. and {Parks}, George K.},
        title = "{Ponderomotive acceleration of ions by circularly polarized electromagnetic waves}",
      journal = {\grl},
         year = 1996,
        month = feb,
       volume = {23},
       number = {4},
        pages = {327-330},
          doi = {10.1029/96GL00157},
       adsurl = {https://ui.adsabs.harvard.edu/abs/1996GeoRL..23..327L}
}

@ARTICLE{1963SPhD....7..988G,
       author = {{Galeev}, A.~A. and {Oraevskii}, V.~N.},
        title = "{The Stability of Alfv{\'e}n Waves}",
      journal = {Soviet Physics Doklady},
         year = 1963,
        month = may,
       volume = {7},
        pages = {988},
       adsurl = {https://ui.adsabs.harvard.edu/abs/1963SPhD....7..988G}
}

@ARTICLE{1966PhFl....9.1483B,
       author = {{Barnes}, Aaron},
        title = "{Collisionless Damping of Hydromagnetic Waves}",
      journal = {Physics of Fluids},
         year = 1966,
        month = aug,
       volume = {9},
       number = {8},
        pages = {1483-1495},
          doi = {10.1063/1.1761882},
       adsurl = {https://ui.adsabs.harvard.edu/abs/1966PhFl....9.1483B}
}

@ARTICLE{1972JPlPh...8..197B,
       author = {{Barnes}, Aaron and {Hung}, R.~J.},
        title = "{Plasma heating and acceleration due to Landau damping of hydromagnetic waves}",
      journal = {Journal of Plasma Physics},
         year = 1972,
        month = oct,
       volume = {8},
       number = {2},
        pages = {197-215},
          doi = {10.1017/S002237780000708X},
       adsurl = {https://ui.adsabs.harvard.edu/abs/1972JPlPh...8..197B}
}

@ARTICLE{1978ApJ...224.1013D,
       author = {{Derby}, N.~F., Jr.},
        title = "{Modulational instability of finite-amplitude, circularly polarized Alfv{\'e}n waves.}",
      journal = {\apj},
         year = 1978,
        month = sep,
       volume = {224},
        pages = {1013-1016},
          doi = {10.1086/156451},
       adsurl = {https://ui.adsabs.harvard.edu/abs/1978ApJ...224.1013D}
}

@ARTICLE{1990JGR....9510525I,
       author = {{Inhester}, B.},
        title = "{A drift-kinetic treatment of the parametric decay of large-amplitude Alfven waves}",
      journal = {\jgr},
         year = 1990,
        month = jul,
       volume = {95},
        pages = {10525-10539},
          doi = {10.1029/JA095iA07p10525},
       adsurl = {https://ui.adsabs.harvard.edu/abs/1990JGR....9510525I}
}

@ARTICLE{1993JGR....9813247J,
       author = {{Jayanti}, Venku and {Hollweg}, Joseph V.},
        title = "{On the dispersion relations for parametric instabilities of parallel-progagating Alfv{\'e}n waves}",
      journal = {\jgr},
         year = 1993,
        month = aug,
       volume = {98},
       number = {A8},
        pages = {13247-13252},
          doi = {10.1029/93JA00920},
       adsurl = {https://ui.adsabs.harvard.edu/abs/1993JGR....9813247J}
}

@ARTICLE{1994JGR....9923431H,
       author = {{Hollweg}, Joseph V.},
        title = "{Beat, modulational, and decay instabilities of a circularly polarized Alfv{\'e}n wave}",
      journal = {\jgr},
         year = 1994,
        month = dec,
       volume = {99},
       number = {A12},
        pages = {23431-23448},
          doi = {10.1029/94JA02185},
       adsurl = {https://ui.adsabs.harvard.edu/abs/1994JGR....9923431H}
}

@ARTICLE{2001A&A...367..705D,
       author = {{Del Zanna}, L. and {Velli}, M. and {Londrillo}, P.},
        title = "{Parametric decay of circularly polarized Alfv{\'e}n waves: Multidimensional simulations in periodic and open domains}",
      journal = {\aap},
         year = 2001,
        month = feb,
       volume = {367},
        pages = {705-718},
          doi = {10.1051/0004-6361:20000455},
       adsurl = {https://ui.adsabs.harvard.edu/abs/2001A&A...367..705D}
}

@ARTICLE{2006PhPl...13l4501N,
       author = {{Nariyuki}, Y. and {Hada}, T.},
        title = "{Kinetically modified parametric instabilities of circularly polarized Alfv{\'e}n waves: Ion kinetic effects}",
      journal = {Physics of Plasmas},
         year = 2006,
        month = dec,
       volume = {13},
       number = {12},
          eid = {124501},
        pages = {124501},
          doi = {10.1063/1.2399468},
archivePrefix = {arXiv},
       eprint = {physics/0608306},
 primaryClass = {physics.plasm-ph},
       adsurl = {https://ui.adsabs.harvard.edu/abs/2006PhPl...13l4501N}
}

@ARTICLE{2015JPlPh..81a3202D,
       author = {{Del Zanna}, L. and {Matteini}, L. and {Landi}, S. and {Verdini}, A. and {Velli}, M.},
        title = "{Parametric decay of parallel and oblique Alfv{\'e}n waves in the expanding solar wind}",
      journal = {Journal of Plasma Physics},
         year = 2015,
        month = jan,
       volume = {81},
       number = {1},
          eid = {325810102},
        pages = {325810102},
          doi = {10.1017/S0022377814000579},
archivePrefix = {arXiv},
       eprint = {1407.5851},
 primaryClass = {astro-ph.SR},
       adsurl = {https://ui.adsabs.harvard.edu/abs/2015JPlPh..81a3202D}
}

@ARTICLE{2017ApJ...842...63S,
       author = {{Shi}, Mijie and {Li}, Hui and {Xiao}, Chijie and {Wang}, Xiaogang},
        title = "{The Parametric Decay Instability of Alfv{\'e}n Waves in Turbulent Plasmas and the Applications in the Solar Wind}",
      journal = {\apj},
         year = 2017,
        month = jun,
       volume = {842},
       number = {1},
          eid = {63},
        pages = {63},
          doi = {10.3847/1538-4357/aa71b6},
archivePrefix = {arXiv},
       eprint = {1705.03829},
 primaryClass = {physics.space-ph},
       adsurl = {https://ui.adsabs.harvard.edu/abs/2017ApJ...842...63S}
}

@ARTICLE{2022RvMPP...6...22N,
       author = {{Nariyuki}, Yasuhiro},
        title = "{Low-frequency Alfv{\'e}n waves and parametric instabilities in fluid and kinetic plasmas}",
      journal = {Reviews of Modern Plasma Physics},
         year = 2022,
        month = dec,
       volume = {6},
       number = {1},
          eid = {22},
        pages = {22},
          doi = {10.1007/s41614-022-00085-1},
       adsurl = {https://ui.adsabs.harvard.edu/abs/2022RvMPP...6...22N}
}

@ARTICLE{1998PhRvD..57.3219T,
       author = {{Thompson}, Christopher and {Blaes}, Omer},
        title = "{Magnetohydrodynamics in the extreme relativistic limit}",
      journal = {\prd},
         year = 1998,
        month = mar,
       volume = {57},
       number = {6},
        pages = {3219-3234},
          doi = {10.1103/PhysRevD.57.3219},
       adsurl = {https://ui.adsabs.harvard.edu/abs/1998PhRvD..57.3219T}
}

@ARTICLE{2019ApJ...881...13L,
       author = {{Li}, Xinyu and {Zrake}, Jonathan and {Beloborodov}, Andrei M.},
        title = "{Dissipation of Alfv{\'e}n Waves in Relativistic Magnetospheres of Magnetars}",
      journal = {\apj},
         year = 2019,
        month = aug,
       volume = {881},
       number = {1},
          eid = {13},
        pages = {13},
          doi = {10.3847/1538-4357/ab2a03},
archivePrefix = {arXiv},
       eprint = {1810.10493},
 primaryClass = {astro-ph.HE},
       adsurl = {https://ui.adsabs.harvard.edu/abs/2019ApJ...881...13L}
}

@ARTICLE{2019MNRAS.483.1731L,
       author = {{Lyubarsky}, Yuri},
        title = "{Radio emission of the Crab and Crab-like pulsars}",
      journal = {\mnras},
         year = 2019,
        month = feb,
       volume = {483},
       number = {2},
        pages = {1731-1736},
          doi = {10.1093/mnras/sty3233},
archivePrefix = {arXiv},
       eprint = {1811.11122},
 primaryClass = {astro-ph.HE},
       adsurl = {https://ui.adsabs.harvard.edu/abs/2019MNRAS.483.1731L}
}

@ARTICLE{2023ApJ...957..102G,
       author = {{Golbraikh}, Ephim and {Lyubarsky}, Yuri},
        title = "{On the Escape of Low-frequency Waves from Magnetospheres of Neutron Stars}",
      journal = {\apj},
         year = 2023,
        month = nov,
       volume = {957},
       number = {2},
          eid = {102},
        pages = {102},
          doi = {10.3847/1538-4357/acfa78},
archivePrefix = {arXiv},
       eprint = {2309.09218},
 primaryClass = {astro-ph.HE},
       adsurl = {https://ui.adsabs.harvard.edu/abs/2023ApJ...957..102G}
}

@ARTICLE{1969ApJ...157..869G,
       author = {{Goldreich}, Peter and {Julian}, William H.},
        title = "{Pulsar Electrodynamics}",
      journal = {\apj},
         year = 1969,
        month = aug,
       volume = {157},
        pages = {869},
          doi = {10.1086/150119},
       adsurl = {https://ui.adsabs.harvard.edu/abs/1969ApJ...157..869G}
}

@ARTICLE{1995MNRAS.275..255T,
       author = {{Thompson}, Christopher and {Duncan}, Robert C.},
        title = "{The soft gamma repeaters as very strongly magnetized neutron stars - I. Radiative mechanism for outbursts}",
      journal = {\mnras},
         year = 1995,
        month = jul,
       volume = {275},
       number = {2},
        pages = {255-300},
          doi = {10.1093/mnras/275.2.255},
       adsurl = {https://ui.adsabs.harvard.edu/abs/1995MNRAS.275..255T}
}

@ARTICLE{2020ApJ...904L..15I,
       author = {{Ioka}, Kunihito},
        title = "{Fast Radio Burst Breakouts from Magnetar Burst Fireballs}",
      journal = {\apjl},
         year = 2020,
        month = dec,
       volume = {904},
       number = {2},
          eid = {L15},
        pages = {L15},
          doi = {10.3847/2041-8213/abc6a3},
archivePrefix = {arXiv},
       eprint = {2008.01114},
 primaryClass = {astro-ph.HE},
       adsurl = {https://ui.adsabs.harvard.edu/abs/2020ApJ...904L..15I}
}

@ARTICLE{2021ApJ...906L..12Y,
       author = {{Yang}, Yu-Han and {Zhang}, Bin-Bin and {Lin}, Lin and {Zhang}, Bing and {Zhang}, Guo-Qiang and {Yang}, Yi-Si and {Tu}, Zuo-Lin and {Zou}, Jin-Hang and {Ye}, Hao-Yang and {Wang}, Fa-Yin and {Dai}, Zi-Gao},
        title = "{Bursts before Burst: A Comparative Study on FRB 200428-associated and FRB-absent X-Ray Bursts from SGR J1935+2154}",
      journal = {\apjl},
         year = 2021,
        month = jan,
       volume = {906},
       number = {2},
          eid = {L12},
        pages = {L12},
          doi = {10.3847/2041-8213/abd02a},
archivePrefix = {arXiv},
       eprint = {2009.10342},
 primaryClass = {astro-ph.HE},
       adsurl = {https://ui.adsabs.harvard.edu/abs/2021ApJ...906L..12Y}
}

@ARTICLE{2021NatAs...5..414K,
       author = {{Kirsten}, F. and {Snelders}, M.~P. and {Jenkins}, M. and {Nimmo}, K. and {van den Eijnden}, J. and {Hessels}, J.~W.~T. and {Gawro{\'n}ski}, M.~P. and {Yang}, J.},
        title = "{Detection of two bright radio bursts from magnetar SGR 1935 + 2154}",
      journal = {Nature Astronomy},
         year = 2021,
        month = apr,
       volume = {5},
        pages = {414-422},
          doi = {10.1038/s41550-020-01246-3},
archivePrefix = {arXiv},
       eprint = {2007.05101},
 primaryClass = {astro-ph.HE},
       adsurl = {https://ui.adsabs.harvard.edu/abs/2021NatAs...5..414K}
}

@ARTICLE{2023MNRAS.519.4094W,
       author = {{Wada}, Tomoki and {Ioka}, Kunihito},
        title = "{Expanding fireball in magnetar bursts and fast radio bursts}",
      journal = {\mnras},
         year = 2023,
        month = mar,
       volume = {519},
       number = {3},
        pages = {4094-4109},
          doi = {10.1093/mnras/stac3681},
archivePrefix = {arXiv},
       eprint = {2208.14320},
 primaryClass = {astro-ph.HE},
       adsurl = {https://ui.adsabs.harvard.edu/abs/2023MNRAS.519.4094W}
}

@ARTICLE{2026arXiv260101169K,
       author = {{Kamijima}, Shoma F. and {Nishiura}, Rei and {Iwamoto}, Masanori and {Ioka}, Kunihito},
        title = "{One-dimensional PIC Simulation of Induced Compton Scattering in Magnetized Electron-Positron Pair Plasma}",
      journal = {arXiv e-prints},
         year = 2026,
        month = jan,
          eid = {arXiv:2601.01169},
        pages = {arXiv:2601.01169},
          doi = {10.48550/arXiv.2601.01169},
archivePrefix = {arXiv},
       eprint = {2601.01169},
 primaryClass = {astro-ph.HE},
       adsurl = {https://ui.adsabs.harvard.edu/abs/2026arXiv260101169K}
}

@ARTICLE{1975Ap&SS..36..303B,
       author = {{Blandford}, R.~D. and {Scharlemann}, E.~T.},
        title = "{On Induced Compton Scattering by Relativistic Particles}",
      journal = {\apss},
         year = 1975,
        month = sep,
       volume = {36},
       number = {2},
        pages = {303-317},
          doi = {10.1007/BF00645256},
       adsurl = {https://ui.adsabs.harvard.edu/abs/1975Ap&SS..36..303B}
}

@ARTICLE{1996A&A...312..937L,
       author = {{Lipunov}, V.~M. and {Panchenko}, I.~E.},
        title = "{Pulsars revived by gravitational waves.}",
      journal = {\aap},
         year = 1996,
        month = aug,
       volume = {312},
        pages = {937-940},
          doi = {10.48550/arXiv.astro-ph/9608155},
archivePrefix = {arXiv},
       eprint = {astro-ph/9608155},
 primaryClass = {astro-ph},
       adsurl = {https://ui.adsabs.harvard.edu/abs/1996A&A...312..937L}
}

@ARTICLE{2001MNRAS.322..695H,
       author = {{Hansen}, Brad M.~S. and {Lyutikov}, Maxim},
        title = "{Radio and X-ray signatures of merging neutron stars}",
      journal = {\mnras},
         year = 2001,
        month = apr,
       volume = {322},
       number = {4},
        pages = {695-701},
          doi = {10.1046/j.1365-8711.2001.04103.x},
archivePrefix = {arXiv},
       eprint = {astro-ph/0003218},
 primaryClass = {astro-ph},
       adsurl = {https://ui.adsabs.harvard.edu/abs/2001MNRAS.322..695H}
}

@ARTICLE{2020PTEP.2020j3E01W,
       author = {{Wada}, Tomoki and {Shibata}, Masaru and {Ioka}, Kunihito},
        title = "{Analytic properties of the electromagnetic field of binary compact stars and electromagnetic precursors to gravitational waves}",
      journal = {Progress of Theoretical and Experimental Physics},
         year = 2020,
        month = oct,
       volume = {2020},
       number = {10},
          eid = {103E01},
        pages = {103E01},
          doi = {10.1093/ptep/ptaa126},
archivePrefix = {arXiv},
       eprint = {2008.04661},
 primaryClass = {astro-ph.HE},
       adsurl = {https://ui.adsabs.harvard.edu/abs/2020PTEP.2020j3E01W}
}

@ARTICLE{2000ApJ...537..327I,
       author = {{Ioka}, Kunihito and {Taniguchi}, Keisuke},
        title = "{Gravitational Waves from Inspiraling Compact Binaries with Magnetic Dipole Moments}",
      journal = {\apj},
         year = 2000,
        month = jul,
       volume = {537},
       number = {1},
        pages = {327-333},
          doi = {10.1086/309004},
archivePrefix = {arXiv},
       eprint = {astro-ph/0001218},
 primaryClass = {astro-ph},
       adsurl = {https://ui.adsabs.harvard.edu/abs/2000ApJ...537..327I}
}

@ARTICLE{2021NatAs...5..378L,
       author = {{Li}, C.~K. and {Lin}, L. and {Xiong}, S.~L. and {Ge}, M.~Y. and {Li}, X.~B. and {Li}, T.~P. and {Lu}, F.~J. and {Zhang}, S.~N. and {Tuo}, Y.~L. and {Nang}, Y. and {Zhang}, B. and {Xiao}, S. and {Chen}, Y. and {Song}, L.~M. and {Xu}, Y.~P. and {Liu}, C.~Z. and {Jia}, S.~M. and {Cao}, X.~L. and {Qu}, J.~L. and {Zhang}, S. and {Gu}, Y.~D. and {Liao}, J.~Y. and {Zhao}, X.~F. and {Tan}, Y. and {Nie}, J.~Y. and {Zhao}, H.~S. and {Zheng}, S.~J. and {Zheng}, Y.~G. and {Luo}, Q. and {Cai}, C. and {Li}, B. and {Xue}, W.~C. and {Bu}, Q.~C. and {Chang}, Z. and {Chen}, G. and {Chen}, L. and {Chen}, T.~X. and {Chen}, Y.~B. and {Chen}, Y.~P. and {Cui}, W. and {Cui}, W.~W. and {Deng}, J.~K. and {Dong}, Y.~W. and {Du}, Y.~Y. and {Fu}, M.~X. and {Gao}, G.~H. and {Gao}, H. and {Gao}, M. and {Gu}, Y.~D. and {Guan}, J. and {Guo}, C.~C. and {Han}, D.~W. and {Huang}, Y. and {Huo}, J. and {Jiang}, L.~H. and {Jiang}, W.~C. and {Jin}, J. and {Jin}, Y.~J. and {Kong}, L.~D. and {Li}, G. and {Li}, M.~S. and {Li}, W. and {Li}, X. and {Li}, X.~F. and {Li}, Y.~G. and {Li}, Z.~W. and {Liang}, X.~H. and {Liu}, B.~S. and {Liu}, G.~Q. and {Liu}, H.~W. and {Liu}, X.~J. and {Liu}, Y.~N. and {Lu}, B. and {Lu}, X.~F. and {Luo}, T. and {Ma}, X. and {Meng}, B. and {Ou}, G. and {Sai}, N. and {Shang}, R.~C. and {Song}, X.~Y. and {Sun}, L. and {Tao}, L. and {Wang}, C. and {Wang}, G.~F. and {Wang}, J. and {Wang}, W.~S. and {Wang}, Y.~S. and {Wen}, X.~Y. and {Wu}, B.~B. and {Wu}, B.~Y. and {Wu}, M. and {Xiao}, G.~C. and {Xu}, H. and {Yang}, J.~W. and {Yang}, S. and {Yang}, Y.~J. and {Yang}, Yi-Jung and {Yi}, Q.~B. and {Yin}, Q.~Q. and {You}, Y. and {Zhang}, A.~M. and {Zhang}, C.~M. and {Zhang}, F. and {Zhang}, H.~M. and {Zhang}, J. and {Zhang}, T. and {Zhang}, W. and {Zhang}, W.~C. and {Zhang}, W.~Z. and {Zhang}, Y. and {Zhang}, Yue and {Zhang}, Y.~F. and {Zhang}, Y.~J. and {Zhang}, Z. and {Zhang}, Zhi and {Zhang}, Z.~L. and {Zhou}, D.~K. and {Zhou}, J.~F. and {Zhu}, Y. and {Zhu}, Y.~X. and {Zhuang}, R.~L.},
        title = "{HXMT identification of a non-thermal X-ray burst from SGR J1935+2154 and with FRB 200428}",
      journal = {Nature Astronomy},
         year = 2021,
        month = apr,
       volume = {5},
        pages = {378-384},
          doi = {10.1038/s41550-021-01302-6},
archivePrefix = {arXiv},
       eprint = {2005.11071},
 primaryClass = {astro-ph.HE},
       adsurl = {https://ui.adsabs.harvard.edu/abs/2021NatAs...5..378L}
}

\end{document}